\documentclass[twocolumn]{aastex631}
\usepackage{color}
\usepackage{multirow}
\usepackage{amsmath}
\usepackage{booktabs}
\newcommand{\elgordoFULL}{ACT-CL~J0102-4915}
\newcommand{\elgordo}{El Gordo}

\newcommand{\LCDM}{$\Lambda$CDM}
\newcommand{\HST}{{\it HST}}
\newcommand{\mytilde}{\raise.19ex\hbox{$\scriptstyle\sim$}}

\newcommand{\kms}{$\mbox{km}~\mbox{s}^{-1}$}

\newcommand{\solarm}{$10^{14}~M_{\sun}$}

\newcommand{\solarmA}{$10^{15}~M_{\sun}$}

\newcommand{\persqarcmin}{arcmin$^{-2}$}
\newcommand{\sqarcmin}{arcmin$^{2}$}

\shorttitle{WL Study of El Gordo}
\shortauthors{Kim et al.}

\begin{document}

\title{Head-to-Toe Measurement of El Gordo: Improved Analysis of the Galaxy Cluster ACT-CL~J0102-4915 with New Wide-field {\it Hubble Space Telescope} Imaging Data}

\author{Jinhyub Kim}
\affiliation{Department of Astronomy, Yonsei University, 50 Yonsei-ro, Seoul 03722, Korea; \texttt{jinhyub@yonsei.ac.kr; mkjee@yonsei.ac.kr}}

\author{M. James Jee}
\affiliation{Department of Astronomy, Yonsei University, 50 Yonsei-ro, Seoul 03722, Korea; \texttt{jinhyub@yonsei.ac.kr; mkjee@yonsei.ac.kr}}
\affiliation{Department of Physics, University of California, Davis, One Shields Avenue, Davis, CA 95616, USA}

\author{John~P.~Hughes}
\affiliation{Department of Physics and Astronomy, Rutgers, the State University of New Jersey, 136 Frelinghuysen Road, Piscataway, NJ 08854-8019, USA}

\author{Mijin Yoon}
\affiliation{Ruhr University Bochum, Faculty of Physics and Astronomy, Astronomical Institute (AIRUB), German Centre for Cosmological Lensing, 44780 Bochum, Germany}
\affiliation{Department of Astronomy, Yonsei University, 50 Yonsei-ro, Seoul 03722, Korea; \texttt{jinhyub@yonsei.ac.kr; mkjee@yonsei.ac.kr}}

\author{Kim HyeongHan}
\affiliation{Department of Astronomy, Yonsei University, 50 Yonsei-ro, Seoul 03722, Korea; \texttt{jinhyub@yonsei.ac.kr; mkjee@yonsei.ac.kr}}

\author{Felipe Menanteau}
\affiliation{Center for AstroPhysical Surveys, National Center for Supercomputing Applications, University of Illinois at Urbana-Champaign, Urbana, IL, 61801, USA}
\affiliation{Department of Astronomy, University of Illinois at Urbana-Champaign, 1002 W. Green Street, Urbana, IL 61801, USA}

\author{Crist{\'o}bal Sif{\'o}n}
\affiliation{Instituto de F{\'i}sica, Pontificia Universidad Cat{\'o}lica de Valpara{\'i}so, Casilla 4059, Valpara{\'i}so, Chile}

\author{Luke Hovey}
\affiliation{Theoretical Design Division, Los Alamos National Laboratory, Los Alamos NM 87545, USA}

\author{Prasiddha Arunachalam}
\affiliation{Department of Physics and Astronomy, Rutgers, the State University of New Jersey, 136 Frelinghuysen Road, Piscataway, NJ 08854-8019, USA}

\begin{abstract}

We present an improved weak-lensing (WL) study of the high$-z$ $(z=0.87)$ merging galaxy cluster ACT-CL J0102-4915 (``El Gordo") based on new wide-field \emph{Hubble Space Telescope} (\HST) imaging data. The new imaging data cover the $\mytilde3.5\times\mytilde3.5$ Mpc region centered on the cluster and enable us to detect WL signals beyond the virial radius, which was not possible in previous studies. 
We confirm the binary mass structure consisting of the northwestern (NW) and southeastern (SE) subclusters and the $\mytilde2\sigma$ dissociation between the SE mass peak and the X-ray cool core. We obtain the mass estimates of the subclusters by simultaneously fitting two Navarro-Frenk-White (NFW) halos without employing mass-concentration relations. The masses are $M_{200c}^{NW} = 9.9^{+2.1}_{-2.2}~\times$~\solarm~and $M_{200c}^{SE} = 6.5^{+1.9}_{-1.4}~\times$~\solarm~for the NW and SE subclusters, respectively. The mass ratio is consistent with our previous WL study but significantly different from the previous strong lensing results. This discrepancy is attributed to the use of extrapolation in strong lensing studies because the SE component possesses a higher concentration. 
By superposing the two best-fit NFW halos, we determine the total mass of El Gordo to be $M_{200c} = 2.13^{+0.25}_{-0.23}~\times$~\solarmA, which is $\mytilde23$\% lower than our previous WL result [$M_{200c} =(2.76\pm0.51)~\times$~\solarmA]. Our updated mass is a more direct measurement since we are not extrapolating to $R_{200c}$ as in all previous studies. The new mass is compatible with the current \LCDM~cosmology. 
\end{abstract}




\section{Introduction}
Studies of galaxy clusters, the largest gravitationally bound systems in the universe, are crucial in cosmology and astrophysics. 
Statistical analysis of the average properties of large samples of clusters enables us to understand the formation and evolution of large-scale structures, constrain cosmology, and test our hierarchical structure formation paradigm (see \citealt{KB2012} for review). 
On the other hand, individual clusters provide the ideal laboratory for multi-wavelength studies of many important astrophysical processes such as galaxy-cluster interaction, plasma turbulence, magnetic field evolution, particle acceleration, dark matter self-interaction, etc (e.g., \citealt{Carilli2002}; \citealt{Clarke2006}; \citealt{Clowe2006}; \citealt{Markevitch2007}; \citealt{Jee2015}; \citealt{Stroe2015}; \citealt{vanWeeren2017}; \citealt{Rajpurohit2018}; \citealt{Ichinohe2019}; \citealt{Rojas2021}).

There are rare populations of clusters with high scientific leverages for cosmology and astrophysics. 
For example, massive high-redshift clusters serve as a powerful test of the current \LCDM~cosmology since the very high end of the high-z cluster mass function is extremely sensitive to cosmological parameters (e.g., \citealt{Allen2011}). 
Radio relic clusters are examples of high-leverage clusters for astrophysics (see \citealt{vanWeeren2019} for review). Since the radio relics are believed to trace the merger shocks, their existence, geometry, and spectral properties provide strong constraints on the phase and configuration of the mergers (e.g., \citealt{Andrade-Santos2019}; \citealt{Wilber2019}). 
Even rarer (thus more useful) are the symmetric double radio relic clusters (e.g., \citealt{Barrena2009}) with a clear bimodality in substructure. These merging clusters are believed to have undergone a relatively simple major merger. In this case, as the two relics probe the same merger, they provide tight constraints on the merging scenario.

This study presents a new weak-lensing (WL) analysis of the massive high-redshift double radio relic cluster \elgordoFULL~based on our wide-field \emph{Hubble Space Telescope} (\HST) imaging data. 
The cluster, more commonly referred to as its nickname ``El Gordo", is a remarkable object with an extremely high mass for its redshfit $z=0.87$ and also symmetric double radio relics. Moreover, its characteristic two cometary tails in X-ray have made the cluster a frequent subject of detailed numerical studies.

\elgordo~was discovered by the Atacama Cosmology Telescope (ACT: \citealt{Menanteau2010}; \citealt{Marriage2011}) as the strongest Sunyaev-Zel'dovich (SZ) decrement in the survey \citep{Hasselfield2013}. 
The optical and X-ray follow-up study (\citealt{Menanteau2012}, hereafter M12) finds that the cluster is indeed extremely massive ($\gtrsim2~\times$~\solarmA) from the velocity dispersion of $\sigma_v=1321\pm106$~\kms~and the intracluster medium (ICM) temperature of $T_{X} = 14.5\pm0.1~\mbox{keV}$. 
Strong lensing (SL) and weak lensing (WL) analyses support this extreme mass (e.g., \citealt{Zitrin2013}; \citealt{Jee2014}; \citealt{Schrabback2018}; \citealt{Diego2020}).

Such an extreme mass of \elgordo~has stimulated a number of discussions on whether or not the presence of the cluster is in tension with the \LCDM~paradigm (e.g., \citealt{Harrison2012}; \citealt{Katz2013}). 
M12 and \cite{Jee2014} (hereafter J14) perform the exclusion curve test \citep{Mortonson2011} based on their mass measurements and claim that the cluster is rare, but compatible with \LCDM~when the full sky is considered. 
When the early ACT+SPT survey area (2800 sq. deg) is assumed for the rarity computation, the previous WL mass lies above the 95\% exclusion curve, although the use of this smaller parent survey volume may be disputable in quantifying the tension.

One of the most critical issues in the above discussion of the tension is the accuracy in mass estimation. 
The mass estimates from the SZ decrement, velocity dispersion, and X-ray data that M12 use are obtained under the hydrostatic equilibrium or scaling relation assumptions, which can be biased for a high-$z$ merging cluster. 
Mass estimates from gravitational lensing do not rely on this assumption. 
The previous SL studies are based on uncertain redshifts of multiple images. Also, significant extrapolations beyond their observed field are required to convert the SL results to the total mass. 
Although the WL signal can be detected far beyond the SL regime, the previous WL studies (e.g., J14; \citealt{Schrabback2018}) also have to use extrapolation because the previous \HST~observations did not extend out to the virial radius of the cluster.

\begin{figure}
\centering
\includegraphics[trim= 0cm 0cm 2.6cm 0cm, width=8.5cm]{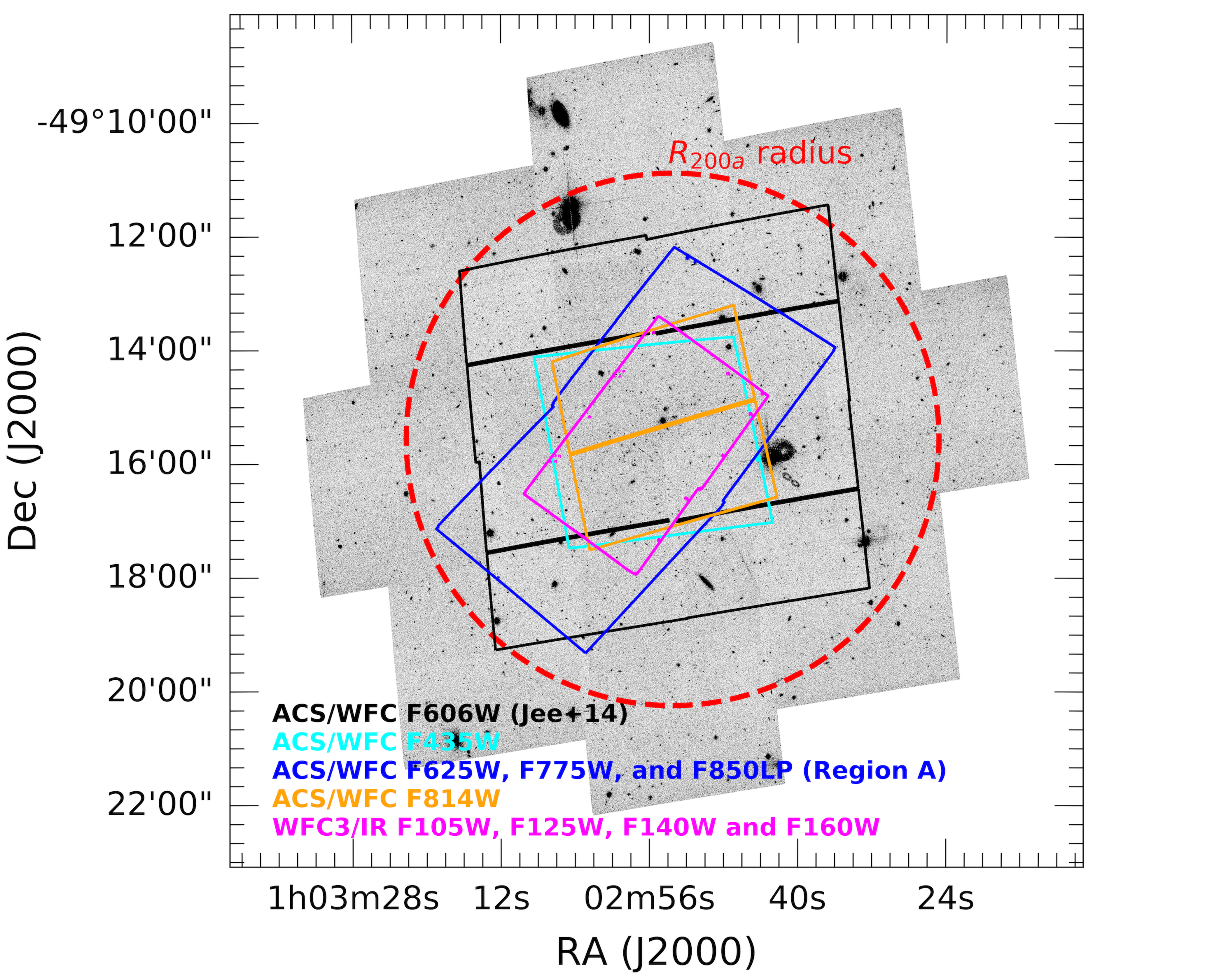}
\caption{\HST~observation footprint of \elgordo. The inverted gray scale represents the intensities in ACS/F606W, which covers a total of $\mytilde119$~\sqarcmin. Our new ACS/F606W observation with eight pointings (PROP 14153) surrounds the previous ACS/F606W observation with a $2\times2$ mosaic pattern (black, PROP 12477), which \cite{Jee2014} used for their weak-lensing analysis. The ACS/F625W, ACS/F775W, and ACS/F850LP data, obtained in a $1\times2$ mosaic pattern with each field centered on one of the two subclusters, cover a $\mytilde6\times3$~\sqarcmin~strip (blue). Throughout the paper, we refer to the areas inside and outside this pointing as region A and region B, respectively. The two WFC3/IR pointings (magenta) lie within the region A. The ACS/F435W (cyan) and ACS/F814W (orange) pointings are also indicated. Note that our $R_{200a}$ radius (red dashed line) is enclosed by the new ACS/F606W footprint (see \S\ref{section_results} for details). }
\label{fig:El_Gordo_field}
\end{figure}

In this paper, we present an accurate mass measurement of \elgordo~without extrapolation using new wide-field \HST~observations.
The new \HST~imaging data comprise the flanking fields of the previous $\mytilde45$~\sqarcmin~Advanced Camera for Surveys (ACS) field (Figure~\ref{fig:El_Gordo_field}). 
This makes the total field of view ($\mytilde119$~\sqarcmin) $\mytilde2.7$ times larger and covers the area beyond the virial radius of the cluster. 
We complement the new wide-field ACS observation with the archival ACS and Wide Field Camera 3 (WFC3) infrared (IR) data that were used in the previous weak and strong lensing analyses, respectively.

Many lines of evidence indicate that \elgordo~is undergoing a major merger. 
M12 identify two subclusters (named ``NW" and ``SE" from their spatial locations) and the cometary structure from the optical spectra and the \emph{Chandra} X-ray data, respectively. 
The binary mass structure has been confirmed by strong (\citealt{Zitrin2013}; \citealt{Cerny2018}; \citealt{Diego2020}) and weak lensing (J14; \citealt{Schrabback2018}) analyses. 
Moreover, M12 suggest the presence of a shock (i.e., radio relics), which is confirmed by follow-up radio observations (e.g., \citealt{Lindner2014}; \citealt{Basu2016}).

However, details of \elgordo's merging scenario are still uncertain. 
The key question is the location of the observed phase during the merging sequence. $N$-body/hydrodynamic simulations (e.g., \citealt{Donnert2014}; \citealt{Molnar2015}; \citealt{Zhang2015}) claim that the two subclusters are moving away from each other after the first core passage (i.e., outgoing phase) employing the observed quantities in X-ray and optical spectroscopy (M12), as well as the WL (J14) results. 
It is worth noting that these simulations did not take account of the location of the radio relics, one of the key information to understand the dynamical history of a merging cluster. 
On the other hand, \cite{Ng2015} argue that the two subclusters are likely to be moving towards each other (i.e., in a returning phase) by comparing the observed distances of the relics from the center of mass to the predicted distances with their data-driven Monte Carlo simulation when the J14 masses are adopted. 
In this work, we revisit the phase of the cluster merger utilizing our updated mass results.

This paper is organized as follows. 
In \S\ref{section_observations}, we describe our new and archival \HST~data and their reduction. 
We present our WL analysis including the theoretical background, modeling the point-spread-function, the shape measurement, the source selection and its redshift determination in \S\ref{section_analysis}. 
Our mass reconstruction and the virial masses of subclusters and the whole system are presented in \S\ref{section_results}. 
In \S\ref{section_discussion}, we compare our new mass estimates with the previous studies and discuss the possible source of mass systematics. Using our new accurate mass estimate, we investigate the rarity of the cluster under the \LCDM~paradigm, and probe the phase of merger. 
We summarize our work and present conclusions in \S\ref{section_summary}.

Throughout the paper, $M_{\Delta c}$ (or $M_{\Delta a}$) corresponds to the mass enclosed within a spherical radius, inside which the mean density equals $\Delta$ times the critical (mean) density of the universe at the cluster redshift ($z=0.870$; M12). We employ $\Delta$ values of $200$ and $500$. 
We choose the cosmology published in \cite{Planck2016}. For this cosmology, the angular diameter distance to the cluster is $\mytilde1636$~Mpc, and thus, the plate scale is $\mytilde476$~kpc~$\mbox{arcmin}^{-1}$ at the cluster redshift. 
We use the AB magnitude system corrected for the Milky Way foreground extinction and quoted uncertainties are at the $1\sigma$ ($\mytilde68.3$\%) level.

\begin{table}
\begin{center}
\caption{\HST~observations of \elgordo~\label{table_observations}}
\begin{tabular}{cccc}
\tableline
\tableline
\\
 \colhead{PROP ID} & \colhead{Filter} & \colhead{$5\sigma$ limiting} & \colhead{Field of View} \\
 & & \colhead{magnitude} & \colhead{(arcmin$^2$)} \\ [1.5ex]
\hline
\multirow{2}{*}{12477$^1$} & ACS/WFC F606W & 27.6 & 44.1 \\ [1.0ex]
 & ACS/WFC F814W & 26.9 & 11.4 \\ [1.5ex]
\hline
\multirow{3}{*}{} & ACS/WFC F625W & 27.4 & 22.8 \\ [1.0ex]
12755$^1$ & ACS/WFC F775W & 27.1 & 22.8 \\ [1.0ex]
 & ACS/WFC F850LP & 26.4 & 22.8 \\ [1.5ex]
\hline
\multirow{5}{*}{} & ACS/WFC F435W & 26.7 & 12.0 \\ [1.0ex]
 & WFC3/IR F105W & 27.3 & 9.5 \\ [1.0ex]
14096$^2$ & WFC3/IR F125W & 26.7 & 9.5 \\ [1.0ex]
 & WFC3/IR F140W & 26.8 & 9.5 \\ [1.0ex]
 & WFC3/IR F160W & 27.2 & 9.5 \\ [1.5ex]
\hline
14153$^3$ & ACS/WFC F606W & 27.6 & 74.6 \\ [1.5ex]
\hline
\hline
\tableline
\end{tabular}
\end{center}
\tablecomments{1. The \HST~observations used in our previous WL analysis \citep{Jee2014}. \\
2. New ACS and WFC3/IR observations in the Reionization Lensing Cluster Survey \citep{RELICS}. \\ 3. New wide-field ACS/F606W observation.}
\end{table}

\section{Observations} \label{section_observations}
\subsection{\HST~data} \label{HST_observations}
We use four \HST~programs (Table~\ref{table_observations}). 
Two of them (PROP IDs 12477 and 12755) were used in the previous WL study (J14). 
Program 12477 (Cycle 19, PI: W. High) consists of ACS Wide Field Channel (WFC) F606W and F814W observations, which have different sets of pointings; the F606W observation used a $2\times2$ mosaic pattern whereas the F814W has a single pointing centered on the midpoint between the two subclusters. 
In PROP 12755 (Cycle 19, PI: J. Hughes), three ACS/WFC filters (F625W, F775W, and F850LP) are employed with two pointings in the NW-SE orientation, covering the two subclusters.

The other two programs (PROP IDs 14096 and 14153) that we add in this work include both ACS and WFC3/IR data. 
Program 14096 (Cycle 23, PI: D. Coe) is a \HST~Treasury Program (Reionization Lensing Cluster Survey, RELICS; \citealt{RELICS}) aiming to find high-redshift galaxies in 41 massive strong-lensing clusters. 
\elgordo~is included in the RELICS and observed with a single ACS/WFC F435W and two WFC3/IR pointings on the central region of the cluster. The WFC3/IR data consist of four filters: F105W, F125W, F140W, and F160W.

Program 14153 (Cycle 23, PI: J. Hughes) is comprised of the eight flanking fields surrounding the previous $2\times2$ F606W observations in program 12477. 
This enables us to amply cover the virial radius of \elgordo~by increasing the field size $\mytilde2.7$ times ($\mytilde119$~\sqarcmin) with respect to the one ($\mytilde45$~\sqarcmin) in J14. 
The new F606W fields were observed with nearly the same orientation as the four original central fields and their centers were chosen to avoid any gaps in the resulting mosaic while maximizing the overall area coverage. 
There are, however, small gaps (covering approximately $0.2$~\sqarcmin) within the original $2\times2$ mosaic (from program 12477) about $2^\prime$ from the center of the mosaic.

The information regarding the $5\sigma$ limiting magnitude and field of view for each filter is summarized in Table~\ref{table_observations}. 
Hereafter, we refer to the region covered by F625W, F775W, and F850LP as ``Region A" where both ACS and WFC/IR colors are available. The area outside ``Region A" is referred to as ``Region B", where only F606W data are available. 
Figure~\ref{fig:El_Gordo_field} illustrates the \HST~footprint of our observations and the locations of these two regions.

Our data reduction procedure is as follows. 
We retrieved the {\tt FLC} and {\tt FLT} images for the ACS and WFC3/IR data, respectively, from the Mikulski Archive for Space Telescopes (MAST)\footnote{https://archive.stsci.edu/}. 
The {\tt FLT} images are calibrated, flat-fielded individual exposures whereas the {\tt FLC} images include the correction for the charge transfer efficiency (CTE) degradation in addition to the calibration applied to the {\tt FLT} images.
J14 showed that the residual shape error after the CTE correction is sufficiently smaller than the statistical noise in the cluster lensing analysis. 
Since the WFC3/IR detector is free from CTE degradation, the {\tt FLT} images are used for our WL analysis.

Since our \HST~observations including the new wide-field F606W imaging data cover a large ($\mytilde119$~\sqarcmin) area, careful image alignment is required. 
By refining astrometric solutions based on common astronomical sources in overlapping regions, we aligned individual exposures with respect to reference images. 
The primary reference image was chosen from the ACS/F814W exposures at the cluster center. 
Since the F606W exposures in program 14153 do not overlap this primary reference frame, we used the ACS/F606W exposures in program 12477 as a secondary reference frame. 
The four secondary reference frames serve as a ``ladder" to estimate the distance between each ACS/F606W exposure in program 14153 and the primary reference frame. 
The mean registration errors between the primary reference frame and the other exposures are $\mytilde0.03$ pixel and $\mytilde0.09$ pixel in regions A and B, respectively.

We used the MultiDrizzle \citep{2002multidrizzle} software to remove cosmic rays, subtract sky background, and combine images. 
In order to produce the sharpest point spread function (PSF), we adopted the Lanczos3 drizzle kernel with the {\tt pixfrac} parameter of $1.0$ and the Gaussian kernel with {\tt pixfrac} parameter of $0.7$ for ACS/WFC and WFC3/IR, respectively \citep[][]{Jee2007a,Jee2017}. 
The output pixel scales of both ACS/WFC and WFC3/IR images were set to 0\farcs05~per pixel. 
As a sanity check, we also performed the data reduction using the DrizzlePac \citep{DrizzlePac} package and confirmed that the two outputs are in excellent agreement ($\mytilde0.02$ pixel).

After the mosaic image creation, sources were detected with SExtractor \citep{Bertin1996} in dual-image mode. 
As the dual-image mode requires a detection image, we created the detection image by combining all available images by weighting their contributions according to the weight maps. 
We detected sources by looking for 10 or more connected pixels that are above 1.5 times the sky rms. 
The total number of sources is $\mytilde29,500$ and the averaged raw number density is $\mytilde250$~\persqarcmin~($\mytilde350$~\persqarcmin~and~$\mytilde220$~\persqarcmin~for regions A and B, respectively).

\subsection{Chandra data} \label{Chandra_observations}
We retrieved three archival \emph{Chandra} observations (ObsID: 12258, 14022, 14023: PI: J. Hughes, ACIS-I detectors, VFAINT Mode, 351~ks in total) of \elgordo. 
We generated level-2 event files with the \texttt{chandra\_repro} script and removed flares with the \texttt{deflare} script in the \emph{Chandra} Interactive Analysis of Observations (\texttt{CIAO}; \citealt{CIAO}) software version 4.12 with \texttt{CALDB} version 4.9.2.1. 
We merged the event files with \texttt{merge\_obs} task and obtained the broad-band (0.5-7~keV) exposure-corrected image. 
Point sources were detected using the \texttt{wavdetect} tool and visually inspected for any spurious detections. They were used to verity the relative astrometric calibration between \HST~and \emph{Chandra} imaging data. 
The X-ray image was adaptively smoothed using the \texttt{aconvolve} tool for visualization purpose. 
We performed spectral fitting using the publicly available \texttt{Python} tool \texttt{ClusterPyXT} \citep{Alden2019} to construct an X-ray temperature map. 
The Galactic hydrogen density and the cluster metal abundance were fixed to be $N_H=1.18\times10^{20}~\mbox{cm}^{-2}$~\citep{HI4PI2016} and 0.3 solar, respectively. 
We excluded the above point sources in this spectral analysis.

\section{Analysis} \label{section_analysis}
\subsection{Theoretical Background} \label{theory}
A foreground lens (e.g., a galaxy cluster) creates distortions in the shapes of background galaxies. The primary goal of the cluster WL analysis is to detect such distortions free from instrumental artifacts, identify the substructures, and quantify the masses. 
Mathematically, this distortion of shape is described as 
\begin{equation}
\textbf{A} = \left (\begin{array} {c c} 1 - \kappa - \gamma_1 & -\gamma _2 \\
                      -\gamma_2 & 1 - \kappa + \gamma_1
          \end{array}  \right ), \label{eqn_lens_trans}
\end{equation}
\noindent
where $\gamma$ is an anisotropic shape distortion (called shear) and $\kappa$ is an isotropic shape distortion (called convergence). 
The convergence $\kappa$ is the projected mass density in units of the critical surface density $\Sigma_c$ defined as 
\begin{equation}
\Sigma_c = \frac{c^2}{4 \pi G D_l \beta}, \label{eqn_sigma_c}
\end{equation}
\noindent
where $c$ is the speed of light, $G$ is the gravitational constant, and $D_l$ is the angular diameter distance to the lens. 
The lensing efficiency $\beta$ is defined as $\beta=D_{ls}$/$D_s$, where $D_{ls}$ ($D_s$) is the angular diameter distance between the lens and the source (between the observer and the source).

Equation~\ref{eqn_lens_trans} transforms a circle into an ellipse with an ellipticity $\gamma=\sqrt{\gamma_1^2 +\gamma_2^2}=(a-b)/(a+b)$, where $a$($b$) is the semi-major(minor) axis of the ellipse. 
Using the orientation of the ellipse $\phi$, the two shear components are given by $(\gamma_1,\gamma_2)=(\gamma \cos 2\phi, \gamma \sin 2\phi)$. 
The quantity that we measured directly from the shape of the source galaxy is the reduced shear $g = \gamma / (1 - \kappa)$. 
In theory, the reduced shear can be obtained by averaging the ellipticities $e = (a-b)/(a+b)$ of the source galaxies.

In fact, the average observed ellipticity does not exactly match the reduced shear because of a number of systematic effects. 
Among the most important are model bias and noise bias. 
The former arises from the difference between the real galaxy profile and the assumed model used in the shape measurement. 
The latter is due to the nonlinear relationship between the ellipticity measurement and the pixel noise. 
We describe our procedure to derive the multiplicative correction factor to calibrate out such biases in \S\ref{section_PSF_shape}.

\subsection{Point Spread Function Modeling and Shape Measurement} \label{section_PSF_shape}
In order to extract the WL signal from each source, it is crucial to know the PSF at the source position. 
An incorrect PSF model adds an additional distortion in source shape, which mimics shear. Even a circular PSF dilutes the lensing signal if uncorrected for with an impact particularly devastating for small (thus faint) galaxies. 
Given that such faint galaxies comprise the majority of the source population in the high-redshift cluster lensing study, accurate PSF modeling is one of the key requirements in our WL analysis.

For ground-based WL studies, one constructs the PSF model based on many high signal-to-noise (S/N) stars in each exposure.
However, this approach is not applicable in a {\it HST} WL study because the field of view is too small; in the case of the ACS/F606W exposures in \elgordo, fewer than 15 stars are available per pointing. 
It is well-known that the \HST~PSF varies in time and position with possible correlations with the thermal cycle of the telescope. This makes the same PSF variation patterns repeat and enables us to use the template-based approach demonstrated in \cite{Jee2007a}, which we adopt in the current analysis. 
Readers are referred to \cite{Jee2007a} for details. 
In brief, we construct the PSF model from dense stellar fields (e.g., globular clusters). Following the prescription of \cite{Jee2007a}, we use a principal component analysis to obtain/identify the most significant basis functions. The amplitude (coefficient) for each basis is interpolated across the detector. 
Therefore, each model is capable of describing the PSF across the entire detector at the time of the observation. 
We link our science exposure to the PSF library by matching the properties (quadrupole moments) of the 10-15 high S/N stars in the science exposures to the corresponding model properties. 
The best-matching models of all exposures are properly rotated and stacked with weights according to their contribution to the final mosaic image at the source galaxy position. 
This PSF library approach has been used in our previous WL studies at high-$z$ for both ACS and WFC3/IR shapes (e.g., J14; \citealt{Jee2017}; \citealt{Kim2019}, hereafter K19; \citealt{Finner2020}). 
In the current study, we make the PSF library for all available filters except for F105W, for which we cannot find adequate stellar field observations for library construction. Thus, we do not use the F105W data for shape measurement.

We determine a galaxy shape by fitting a two-dimensional (2D) elliptical Gaussian model. 
The minimization was carried out with the {\tt MPFIT} \citep{MPFIT} package. 
Prior to shape fitting, the 2D elliptical Gaussian model is convolved with the appropriate PSF model at each galaxy position. 
Among the seven parameters of the 2D elliptical Gaussian model, we fix three parameters (the $x-y$ centroid positions and the background level) to the SExtractor measurements; the remaining four free parameters are the normalization, orientation angle ($\phi$), semi-major ($a$) and -minor ($b$) axes. 
The two ellipticity components are obtained from the following relations:
\begin{eqnarray}
e_1&=&e \cos{2 \phi}, \\
e_2&=&e \sin{2 \phi} \label{e1e2}
\end{eqnarray}
\noindent
where $e = \sqrt{e_1^2 + e_2^2}$. 
The ellipticities from different filters if available are combined with weight-averaging. 
The weight is computed from $1/ \delta_e^{2}$, where $\delta_e$ is the ellipticity measurement error.

As mentioned in \S\ref{theory}, there is a difference between the observed ellipticity and the reduced shear. 
In order to correct for the bias, we utilize the results of our WL shear recovery simulation in \cite{Jee2013} and adopt the multiplicative shear calibration factors for ACS and WFC3/IR filters. 
Here we describe the WL simulation briefly. The simulated images of model galaxies are created by applying arbitrary input shears and degraded down to the \HST~observation conditions (e.g., adding ACS/PSFs as well as noise). 
For each galaxy, shape measurement is performed and ellipticities are obtained. 
By comparing the input shears and the output averaged ellipticities, we determine the slope and adopted it as the shear calibration factor. 
The calibration factor is 1.15 for ACS/WFC filters. Readers are referred to \cite{Jee2013} for details of the WL simulation. 
For WFC3/IR filters, the calibration factor is derived by utilizing $\mytilde2700$ common astronomical sources between ACS/WFC and WFC3/IR in the archival data. 
This approach is preferred over running an additional set of WL simulations for WFC3/IR because our understanding of the WFC3/IR detector systematics (inter-pixel capacitance, persistence, undersampling, nonlinearly, etc.) and capability to simulate the effects are incomplete. 
The resulting calibration factor is estimated to be 1.25 for WFC3/IR \citep{Jee2017}. 
We compared the shapes ($e_1$ and $e_2$) in both ACS and WFC3/IR filters and found that the systematic difference is less than $3$\%, which is negligible when compared to the statistical error.

\subsection{Source Selection and Source Redshift Estimation} \label{section_source_selection}
We select 11,742 galaxies as our background source population that satisfy both shape and photometric requirements described in \S\ref{selection_shape} and \S\ref{selection_photo}. 
This corresponds to a number density of $\mytilde124$~\persqarcmin, which is slightly higher than in J14 ($\mytilde90$~\persqarcmin) and significantly ($\mytilde6.2$ times) higher than in \cite{Schrabback2018} ($\mytilde20$~\persqarcmin). 
We stress that our observation ($\mytilde119$~\sqarcmin) covers a much wider area than the previous WL studies ($\mytilde45$~\sqarcmin) of J14 and S18. 
In \S\ref{selection_beta}, we describe how we determine the source redshift and lensing efficiency.

\subsubsection{Shape Requirement} \label{selection_shape}
We follow the procedures for source selection described in J14 and K19. Here we briefly summarize the conditions that the sources should satisfy. 
First, we select sources with a {\tt MPFIT} {\tt STATUS} value of 1 \citep{MPFIT}, which indicates that the fitting converged without any major issues. 
Second, the background sources are required to have an ellipticity measurement error less than 0.25. This excludes sources with very low surface brightness and/or small FWHM values. 
Third, the half-light radii of the background sources are forced to be larger than those of stars. This measure is efficient in removing bright point sources. 
Last, the sources should have semi-minor axes greater than 0.4 pixel. If this condition is not imposed, some faint point sources remain. Also, many sources failing to meet this condition suffer from pixelization artifacts \citep{Jee2013}. 
Even after these conditions are imposed, there still remain some spurious objects (e.g., diffraction spikes around bright stars, clipped objects at the field boundaries, fragmented parts of foreground galaxies, etc.). 
We removed the remaining spurious objects by visual inspection.

\begin{figure}
\centering
\includegraphics[width=8.5cm]{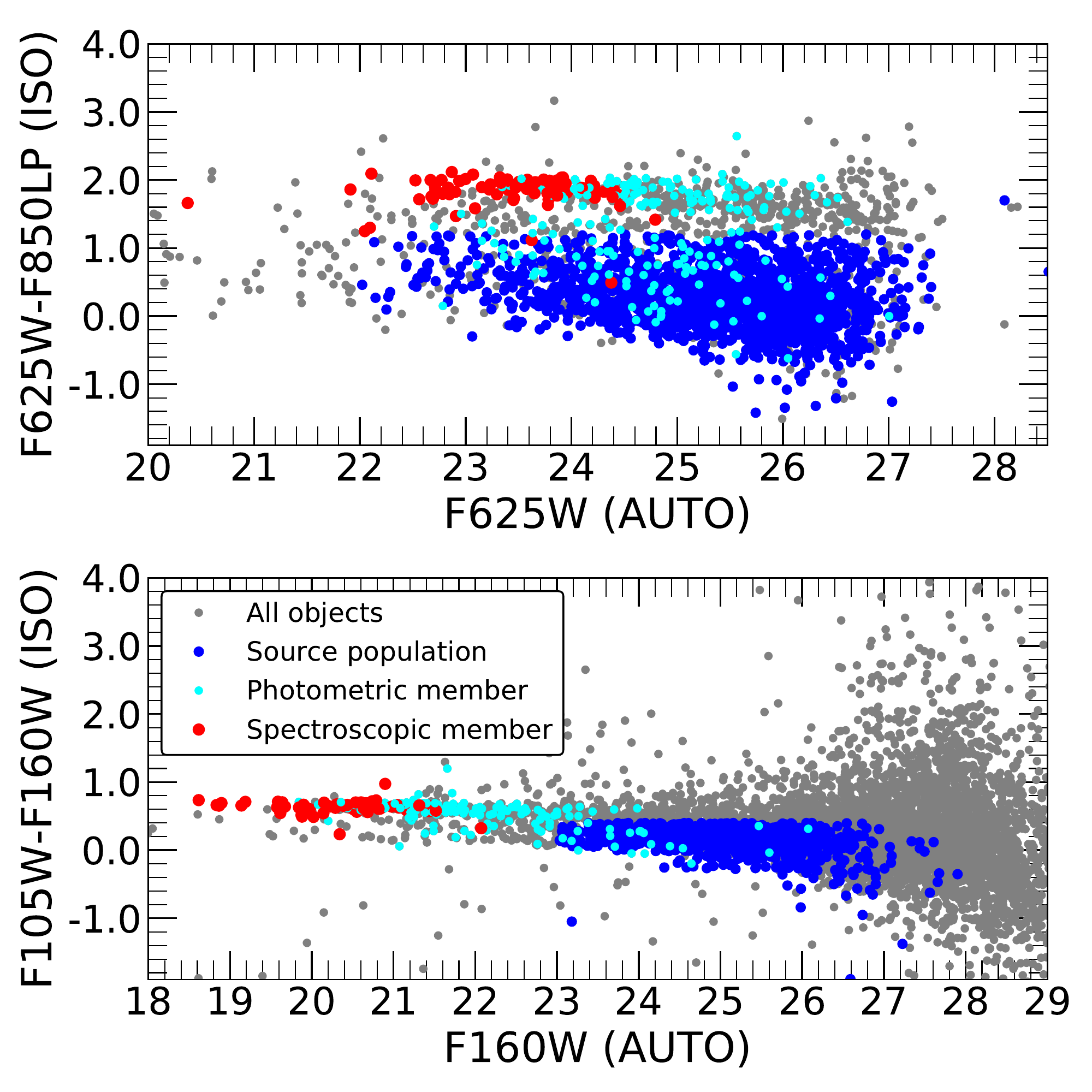}
\caption{Color-magnitude relations in region A. 
The top (bottom) panel shows the result for the central $\mytilde22.8$~\sqarcmin~($\mytilde9.5$~\sqarcmin) region seen in $\mbox{F625W}-\mbox{F850LP}$ versus $\mbox{F625W}$ ($\mbox{F105W}-\mbox{F160W}$ versus $\mbox{F160W}$). 
Sources (blue circles) are selected based on color and magnitude cuts (top: $\mbox{F625W}-\mbox{F850LP}< 1.2$, $22.0 < \mbox{F625W} < 28.0$,
bottom: $\mbox{F105W}-\mbox{F160W}< 0.4$, $23.0 < \mbox{F160W} < 28.0$) and the shape selection criteria (see text). 
The cluster galaxies identified by spectrocopy and photometry in \cite{Menanteau2012} are also indicated by red and cyan, respectively. 
Although most cluster member galaxies tend to be isolated well in $\mbox{F625W}-\mbox{F850LP}$ and in $\mbox{F105W}-\mbox{F160W}$, a number of blue cluster members overlap with the source population in color and thus are removed manually. }
\label{fig:cmd}
\end{figure}

\begin{figure}
\centering
\includegraphics[width=8.5cm]{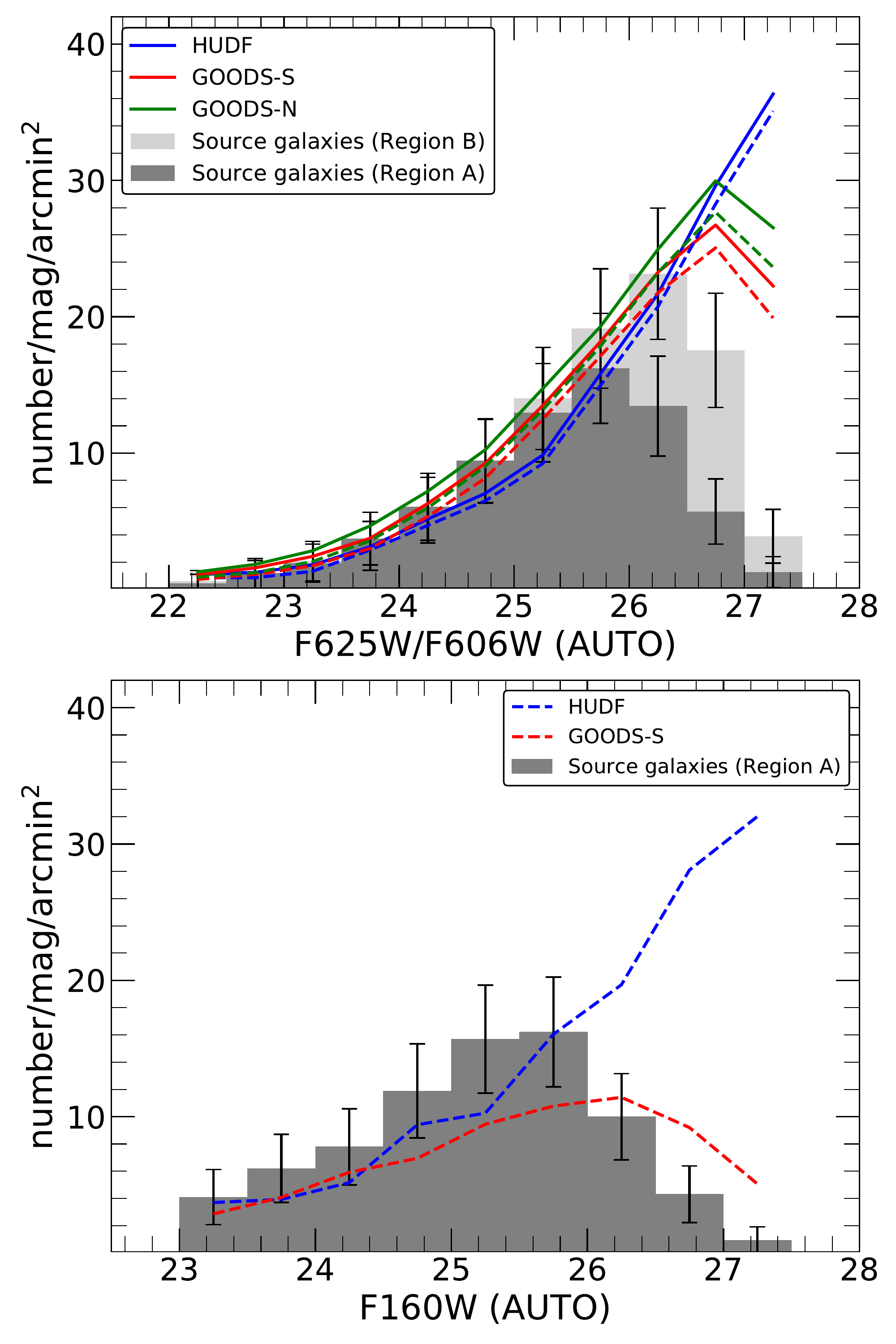}
\caption{Comparison of magnitude distributions of our sources and the objects in the control fields.
The dashed lines represent the control field galaxy magnitude distribution after both color and magnitude cuts are applied whereas the solid lines are the results only with the magnitude cut applied. 
Top: $\mbox{F625W}$ ($\mbox{F606W}$) magnitude comparison between region A (region B) in \elgordo~and the three control fields: HUDF, GOODS-S, and GOODS-N. The source galaxies with (region A) and without color (region B) information are displayed using dark and light gray histograms, respectively. Error bars are computed based on Poissonian statistics. The consistency within the sample variance between our source galaxies and the control field ones shows that the blue member contamination is negligible. 
Bottom: Comparison of the $\mbox{F160W}$ magnitude distribution between \elgordo~and the two control fields: HUDF and GOODS-S. The magnitude distribution of our sources selected with WFC3/IR is consistent with the HUDF one. }
\label{fig:mag_distribution}
\end{figure}

\subsubsection{Photometric Requirement} \label{selection_photo}
Since the 4000~\AA~break feature is redshifted to $\mytilde7500$~\AA~at $z=0.87$, the $\mbox{F625W}-\mbox{F850LP}$ color can be used to identify the cluster red-sequence. 
In the color-magnitude diagram shown in the top panel of Figure~\ref{fig:cmd}, indeed one can readily find the locus of the early-type galaxies. 
We select sources bluer than the red-sequence by applying the photometric selection conditions described in J14 to the galaxies in region A where the ACS color is available: $\mbox{F625W}-\mbox{F850LP}< 1.2$ and $22 < \mbox{F625W} < 28$. 
In region B, since only F606W observations were performed, we rely on the magnitude selection criterion. The galaxies with $22 < \mbox{F606W} < 28$ are selected for our source population.

Potentially, many blue member galaxies could be included into our source galaxies (top panel of Figure~\ref{fig:cmd}), although we remove the known blue cluster members identified by spectroscopic and/or photometric redshift in M12. 
In order to test this potential blue cluster member contamination, we utilize two well-observed control fields: the Hubble Ultra Deep Field (HUDF; \citealt{UVUDF}) and the Great Observatories Origins Deep Survey (GOODS; \citealt{GOODSmag}). 
We impose the same source selection conditions to these control fields and obtain the magnitude distributions (the top panel of Figure~\ref{fig:mag_distribution}). 
If our source galaxies were significantly contaminated by blue members, an excess would be noticeable in this histogram. However, the magnitude distributions of our sources and the ones from the control fields are consistent at $\mbox{F625W} < 26.5$; the difference at $\mbox{F625W} > 26.5$ is due to the depth. 
This implies that the impact of blue member contamination is not significant.

In region B where we only have F606W observation, the contamination from the faint red cluster members and foreground galaxies would be inevitable. We examine the red galaxy contribution by comparing the number density distributions between regions A and B for the same magnitude range. 
The upper panel of Figure~\ref{fig:mag_distribution} shows that the two distributions are consistent at $\mbox{F606W} < 25$, which suggests that the fraction of bright red galaxies is negligible in our source population. 
This consistency is attributed to the removal of red cluster members also in region B based on the spectroscopic and/or photometric redshift in the cluster field (M12). The redshift catalog provides the spec-$z$ and/or photo-$z$ of $415$ member galaxies, which enables us to identify cluster members effectively out to $\mytilde2$ Mpc radius from the center of mass. 
At fainter magnitude ($\mbox{F606W} > 25$), on the other hand, there are indications that the source densities in region B are higher. Given that blue cluster member contamination is negligible and most member galaxies that have either photo-$z$ and spec-$z$ data are bright ($\mbox{F606W} < 25$), such an excess is due to mainly the inclusion of (faint) red galaxies in region B. This faint red population includes foreground and cluster galaxies, which are not lensed and thus dilute the lensing signal. 
In \S\ref{selection_beta}, we determine the source redshift by carefully taking into account such contamination, following the prescription of K19.

In the WFC3/IR photometry, given that the locus of cluster members is well identified by F105W and F160W as seen in the bottom panel of Figure~\ref{fig:cmd}, we select WFC3/IR objects bluer and fainter than the cluster galaxies ($\mbox{F105W}-\mbox{F160W}< 0.4$ and $23 < \mbox{F160W} < 28$). 
We again check for cluster member contamination on the source population. 
The bottom panel of Figure~\ref{fig:mag_distribution} shows that the source number density between our source and the control field is consistent with that from the HUDF at $\mbox{F160W} < 26$. 
The number density of our source population rapidly drops at $\mbox{F160W} > 26$ because our observation is shallower than the HUDF field.

\begin{figure*}
\centering
\includegraphics[trim=1cm 0cm 2.6cm 0cm,width=88mm]{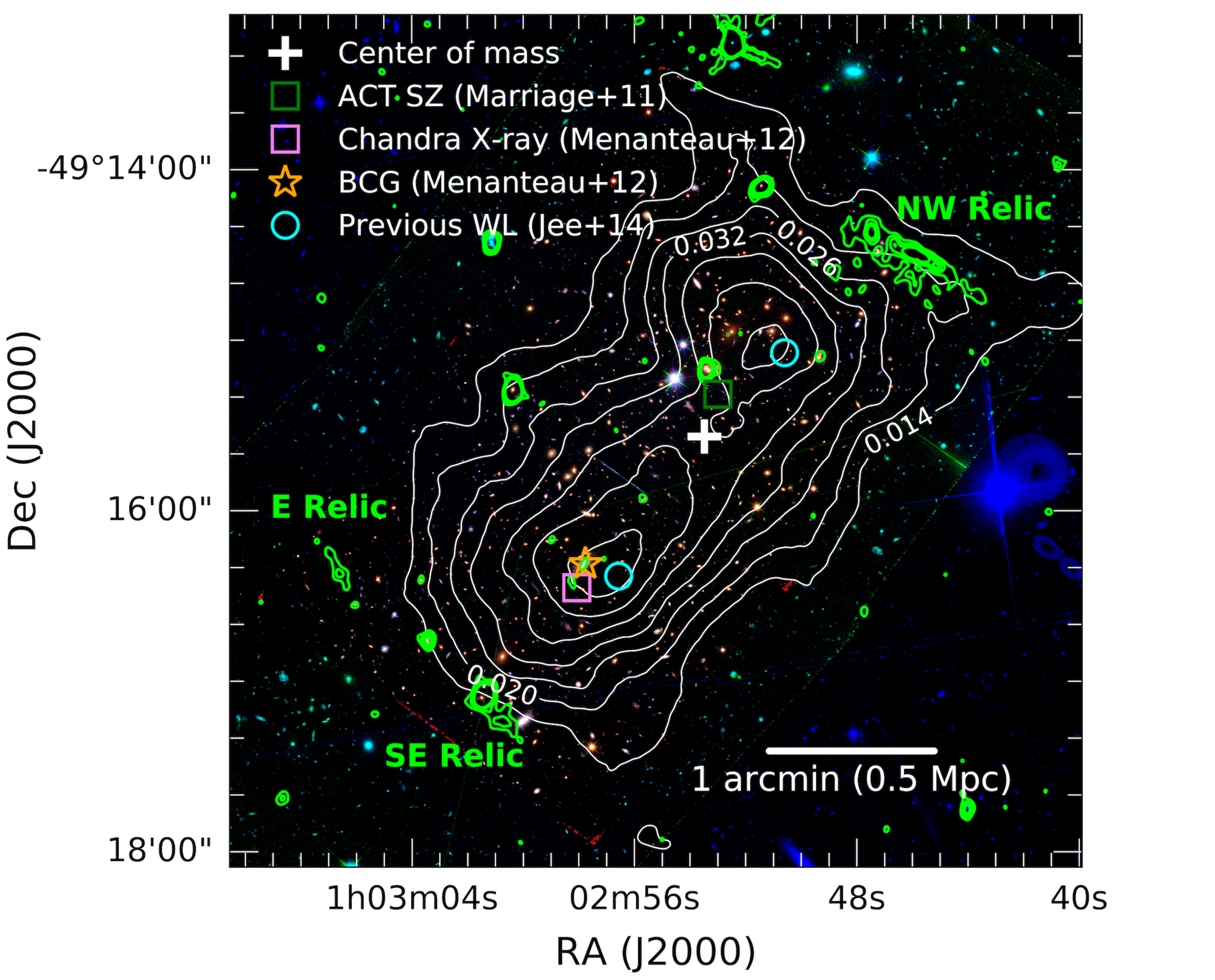}
\includegraphics[trim=1cm 0.5cm 2.6cm 0cm,width=87mm]{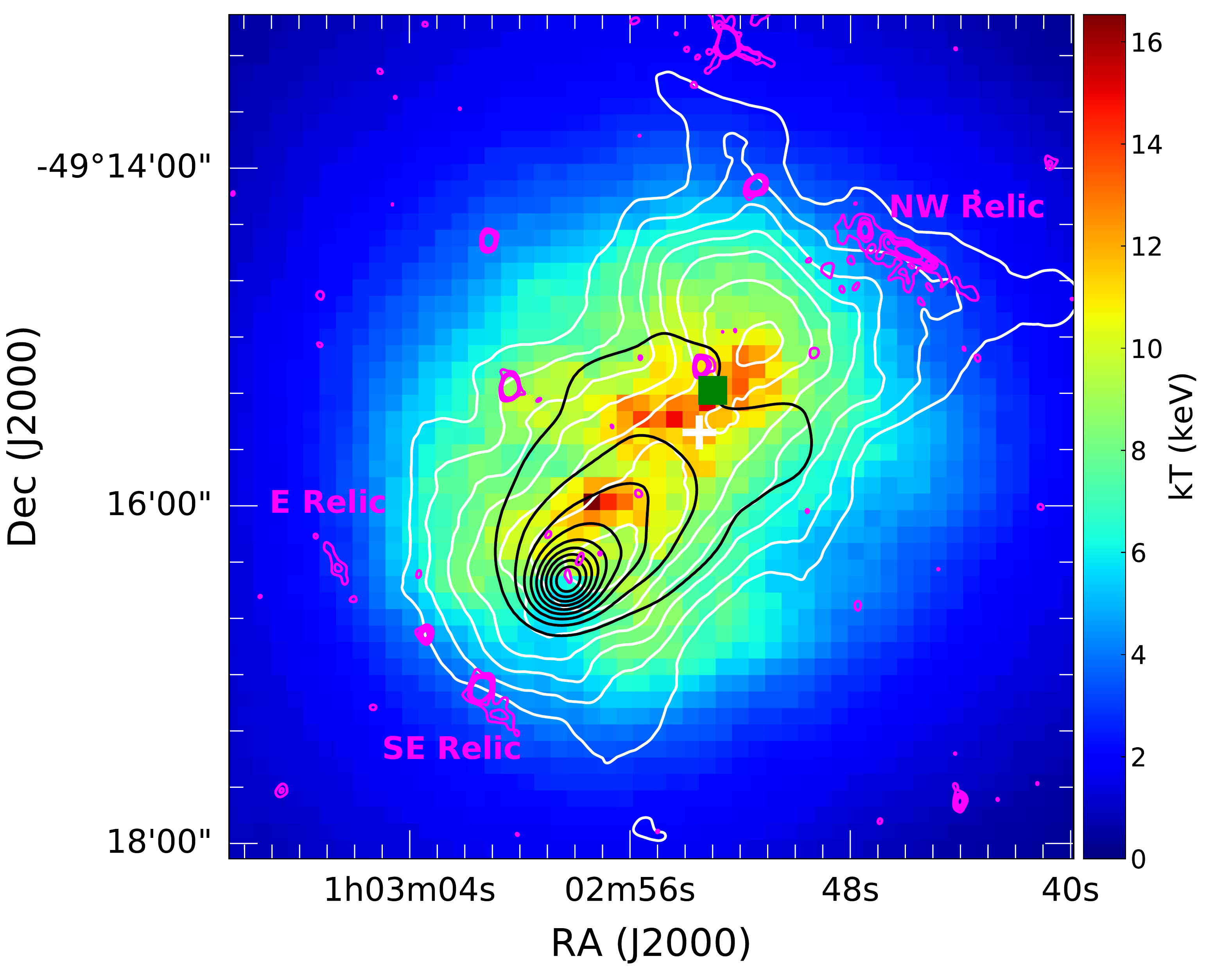}
\caption{Two-dimensional mass reconstruction of \elgordo~with the {\tt FIATMAP} code \citep{FIATMAP}. 
In the left panel, the mass map is overlaid on the color composite created with \HST~ACS/WFC F606W (blue), ACS/WFC F775W (green), and WFC3/IR F105W (red). The symbols represent the centroids measured at various wavelengths as well as the center of mass (white cross). 
The mass contour starts at $2.5\sigma$ significance and the significance increases inward by $1\sigma$, reaching peak significance values of $\mytilde8.8\sigma$ and $\mytilde9.0\sigma$ for the NW and SE peaks, respectively. 
Note that the contours are labeled with values of the the convergence $\kappa$, whose scale is arbitrary because the mass-sheet degeneracy is not broken. 
The right panel shows the same mass contours overlaid on the temperature map created with {\it Chandra} data. The SZ peak (filled green square) coincides with the hot temperature region close to the NW mass peak. The X-ray surface brightness map shown in black reveals only a single peak centered on the SE subcluster. 
In both panels, we also overlay the 2.1 GHz ATCA contours, which display the three radio relics reported in \cite{Lindner2014} as well as some bright radio point sources. The radio contour levels are 5, 10, 15, and 20$\sigma$ ($\sigma=8.2\,\rm\mu Jy\,beam^{-1}$). }
\label{fig:mass_map_HST}
\end{figure*}

\begin{figure*}
\centering
\includegraphics[width=195mm]{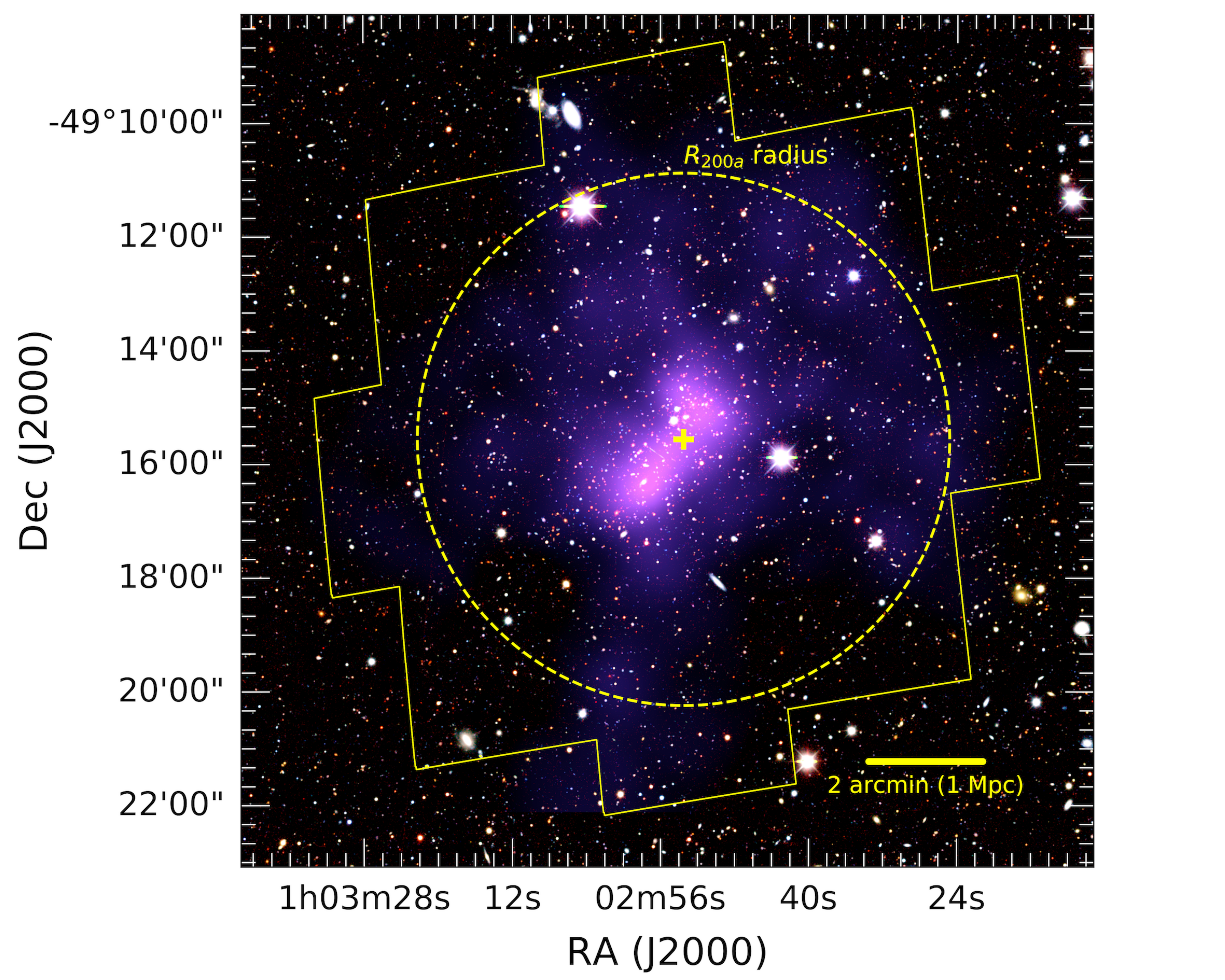}
\caption{Wide-field mass reconstruction of \elgordo.
The intensity in purple represents the convergence obtained with the {\tt MAXENT} code \citep{Jee2007b}. 
Only for demonstration purpose, here we display the $15\arcmin\times15\arcmin$ color composite created by the {\it g}, {\it r}, and {\it z} filters of Dark Energy Camera \citep{DECam} for the blue, green, and red channels, respectively. 
The footprint of our new \HST~observations is indicated with the yellow solid line. 
The $R_{200a}$ radius is marked with the yellow dashed line. 
The center of mass (yellow cross) is computed with the best-fit subcluster masses. }
\label{fig:mass_map_DECam}
\end{figure*}

\subsubsection{Source Redshift Estimation} \label{selection_beta}
Since the lensing efficiency $\beta$ is an important quantity to calculate the surface mass density and thus the WL signal, it is paramount to determine the lensing efficiency of the source population accurately. 
In this work, we obtain this quantity by utilizing the HUDF field after applying the same source selection conditions described in \S\ref{section_source_selection}.

The average lensing efficiency of a source population is determined by
\begin{equation}
\beta = \left < \mbox{max} \left (0,\frac{D_{ls}}{D_s} \right ) \right >. 
\end{equation}
\noindent
We assign $\beta$ to be zero for foreground objects, while obtaining the ratio $D_{ls} / D_{s}$ for background sources. 
Then we take the average lensing efficiency for galaxies within each magnitude bin (magnitude range from A to B with 0.5 magnitude bin size). 
We determine the representative value of the lensing efficiency $\left < \beta \right >$ for the whole source population by taking the weighted mean of the averaged $\beta$ of each magnitude bin. 
In calculating the weights, we address the depth difference between the cluster field and the control field (HUDF in this case) and the inhomogeneity of the number density at each magnitude bin. 
Additionally, we also consider the fraction of red galaxies when color information is unavailable such as the sources in region B.

As K19 pointed out, the source population just selected purely by magnitude includes blue sources, red background/foreground objects, and red cluster galaxies. 
Of course, blue sources alone can only be selected when color information is available. Although the blue sources are composed of blue foreground/background sources as well as blue cluster members, the contamination caused by the blue foreground/cluster galaxies is negligible (see Figure~\ref{fig:mag_distribution} and \textsection\ref{selection_photo}). 
Among the galaxies with red color in the source population, only red background galaxies are lensed. Thus, we should address the impact of the red background galaxies on the lensing efficiency estimation in region B where the sources are selected by magnitude information alone.

We obtain the lensing efficiencies of blue sources and red background galaxies in region B using this equation: 
\begin{equation}
\beta_{regB,i} = \beta_{bs,i}\frac{n_{bs,i}} {n_{regB,i}} + \beta_{rb,i}\frac{n_{rb,i}} {n_{regB,i}} , \label{eqn_beta_regB}
\end{equation}
\noindent 
where $\beta_{bs,i}$ ($n_{bs,i}$) and $\beta_{rb,i}$ ($n_{rb,i}$) are the average lensing efficiencies (number densities) at the i$^{th}$ magnitude bin for blue sources and red background galaxies, respectively. $n_{regB,i}$ is the total number density within each magnitude bin in region B. 
The lensing efficiency of blue sources ($\beta_{bs,i}$) is identical to the $\beta_i$ measured in region A. Also, the number density of blue sources ($n_{bs,i}$) is the same as that measured in region A. 
The number density of red background sources ($n_{rb,i}$) cannot be obtained directly. However, the ratio $n_{rb,i}/n_{regB,i}$ can be inferred from the HUDF photo-$z$ catalog. 
Although the catalog does not include the red cluster member galaxies and thus the inferred ratio can be overestimated, this overestimation is insignificant because the fraction of red cluster members in region B is small (e.g., $\lesssim 5$\% beyond $R_{500c}$ radius). 
For WFC3/IR sources, we do not consider the possibility of the red galaxay inclusion since all WFC3/IR sources have color information.

From the procedure above, the resulting lensing efficiencies are $\left < \beta \right > =0.254$ and $0.220$ (corresponding to the effective redshifts $z_{\rm eff}=1.309$ and $1.226$, respectively) for the ACS sources in region A and B, respectively. 
If we had neglected the impact of the red galaxy inclusion, the lensing efficiency in region B might have increased to $0.272$. 
For the WFC3/IR sources, we obtain $\left < \beta \right >=0.195$ ($z_{\rm eff}=1.172$).

Note that the lensing efficiencies determined above assume that all sources are at the same redshift on average. 
In fact, as the sources must have a distribution of their redshifts, we have to consider the width of the source distribution. \cite{Seitz1997} suggest a first-order correction to the shear, using the lensing efficiency $\left < \beta \right >$ and its width $\left < \beta^{2} \right >$, given by 
\begin{equation}
\frac{g\prime}{g} = 1 + \left
(\frac{\left < \beta^{2} \right >}{\left<\beta\right>^{2}} - 1 \right )\kappa, 
\end{equation}
\noindent
where $g\prime$ and $g$ are the observed and true shears, respectively. 
We obtain the width of the source population by squaring each object's $\beta$ value and then taking the average. The results are $\left < \beta^{2} \right >=0.106$, $0.092$, and $0.076$ for the ACS sources in region A, the ones in region B, and the WFC3/IR sources, respectively.

We investigate the systematic uncertainty due to the sample variance using the test described in J14 and K19. 
We divide the GOOD-S and GOODS-N fields into 9-11 subregions and measure the variation of $\left < \beta \right >$ values. 
Within the GOODS-S and GOODS-N fields, the standard deviations of the $\left < \beta \right >$ distributions are $0.011$ and $0.013$ for ACS and WFC3/IR sources, respectively, which correspond to a $\mytilde5$\% shift in mass.

\begin{figure}
\centering
\includegraphics[width=85mm]{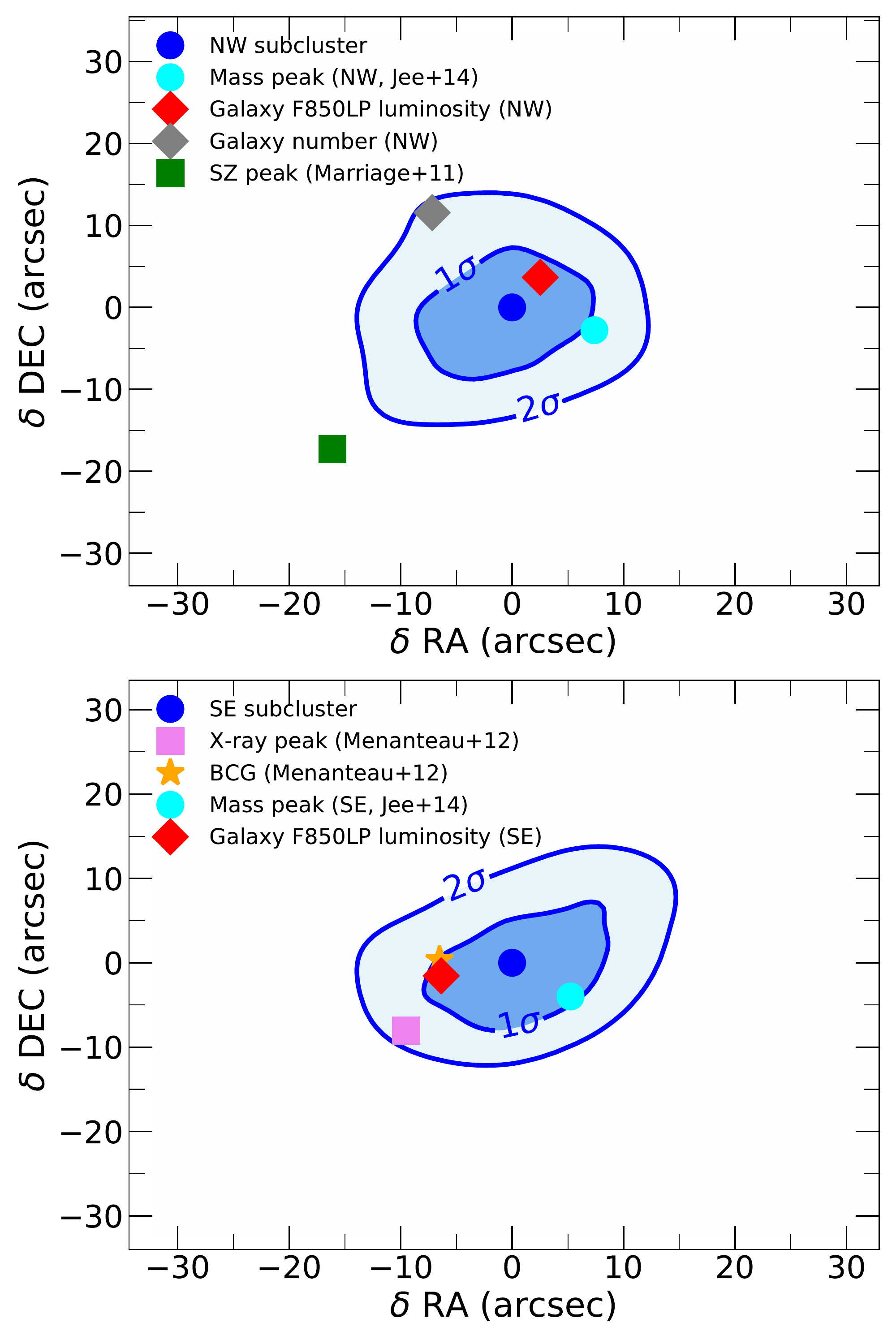}
\caption{Bootstrapping test for the WL mass centroid significance. 
The blue contours are the centroid distributions for the NW (top panel) and SE (bottom panel) in our 1000 bootstrapping runs. The (0,0) positions correspond to the locations of the centroids for the NW component (R.A. = 01:02:51.23, Decl. = -49:15:2.56) and SE one (R.A. = 01:02:56.95, Decl. = -49:16:21.86), respectively. 
The galaxy luminosity and mass peaks from the previous WL work (J14) are located near the $1\sigma$ centroid contours of each subcluster. 
The X-ray peak around the SE component is offset from the SE centroid at the $2\sigma$ level. 
The galaxy number density peaks are also separated from our mass centroids at the $\mytilde2\sigma$ and $\mytilde6.6\sigma$ levels for NW and SE components, respectively. 
The SZ peak is closer to NW with an offset at the $\mytilde3.7\sigma$ level.}
\label{fig:centroids}
\end{figure}

\section{Results} \label{section_results}
\subsection{Mass Reconstruction} \label{mass_map}
Our {\it HST} WL data enable us to obtain a high-fidelity 2D projected mass reconstruction of \elgordo. 
Among various algorithms, we use the {\tt FIATMAP} (\citealt{FIATMAP}) and the maximum entropy ({\tt MAXENT}) reconstruction methods (\citealt{Jee2007b}) for conversion of the ellipticity (reduced shear) field into a mass map. 
The {\tt FIATMAP} mass map (Figure~\ref{fig:mass_map_HST}) is highly consistent with the {\tt MAXENT} result (Figure~\ref{fig:mass_map_DECam}), although the latter provides more efficient noise suppression in the cluster outskirts. 
We refer readers to Appendix~\ref{mass_map_comparison} for side by side comparison. 
For the case of the {\tt FIATMAP} result, we determine the significance of the substructures by creating an uncertainty (rms) map based on 1000 bootstrapping of source galaxies.

We detect the two substructures (hereafter referred to as the NW and SE components) separated by $\mytilde770$~kpc. 
These substructures are consistent with the previous WL results (J14; \citealt{Schrabback2018}). 
In the central region, the significance is higher, reaching nearly $9\sigma$ at each WL peak ($8.8\sigma$ and $9.0\sigma$ for the NW and SE peaks, respectively), thanks to the use of the WFC3/IR images. Also, the new wide-field F606W data reveal the mass structure on a larger scale out to $\mytilde3.5$ Mpc from the cluster center (Figure~\ref{fig:mass_map_DECam}).

One notable change with respect to J14 is the mass density enhancement of the SE peak. 
In J14, the lensing signal for the SE peak is weaker than the one for the NW peak (see Figure 5 of J14). On the other hand, our new mass reconstruction shows that the projected mass density of the SE peak is somewhat higher than the NW peak. 
We attribute the difference to the addition of the WFC/IR filters in the current WL study, which reduces the mass reconstruction noise. 
We confirm that when we repeat our analysis purely based on the ACS imaging data, the density of the SE structure decreases and the resulting mass map more resembles the J14 version.

In fact, a higher mass density for the SE component than the NW one has been suggested by SL studies (e.g., \citealt{Zitrin2013}; \citealt{Cerny2018}; \citealt{Diego2020}). For instance, \cite{Cerny2018} performed SL analysis using six ACS and WFC3/IR filters and reported that the SE component has a $\mytilde20$\% higher mass density (and thus more massive) than the NW one within the SL regime ($\leq500$ kpc). 
The current study based on the new WL data supports the claim that the mass structure of the SE component is more compact. 
However, as we demonstrate in \S\ref{mass_estimates}, we find that the NW cluster is more massive (by a factor of $\mytilde1.5$) when a larger scale is considered. We provide more detailed discussion on the mass ratio in \S\ref{mass_ratio_inconsistency}.

Peak locations from various observations (e.g., X-ray, SZ, member galaxies) probe different cluster components. 
For example, the offset between the mass and X-ray peaks can be used to infer the velocity of the subhalos (J14; \citealt{Ng2015}). 
We examine the statistical uncertainty of our mass centroids and compare them with the peaks from other multi-wavelength observations. 
The centroid uncertainties of our WL mass map are obtained by bootstrapping analysis. 
We determined the centroids of the mass peaks from 1000 mass maps by fitting 2D circular Gaussians to the convergence field. 
Among the fitting parameters, we allow only the centroid $(x, y)$ to be free. We fixed the background, amplitude, and the width of the 2D Gaussian to be zero, the highest convergence value, and the effective smoothing scale (FWHM $\sim17$\arcsec), respectively.

Figure~\ref{fig:centroids} shows the results for both NW and SE mass centroids. 
Both mass centroids coincide with the corresponding galaxy luminosity peaks and the mass centroids of J14 within 1$\sigma$. 
The X-ray peak is $\mytilde12$\arcsec~($\mytilde97$~kpc) offset from the SE mass centroid at the $\mytilde2\sigma$ level. 
Unlike the luminosity peaks, the galaxy number density peaks are offset from their corresponding mass peaks by $\mytilde15$\arcsec~($\mytilde120$~kpc, $\mytilde2\sigma$ level) and $\mytilde45$\arcsec~($\mytilde360$~kpc, $\mytilde6.6\sigma$ level) for the NW and SE mass components, respectively; Figure~\ref{fig:centroids} does not show the SE number density peak because it is outside the plotting range. 
The SZ peak \citep{Marriage2011} is $\mytilde190$~kpc away from the NW mass peak (toward the SE mass peak), and this separation corresponds to $\mytilde3.7\sigma$ level. However, since the beam size of the SZ image is large (1\farcm4) and the gas density and temperature of the cluster are spatially complex, it is difficult to interpret the offset.

\begin{figure}
\centering
\includegraphics[width=85mm]{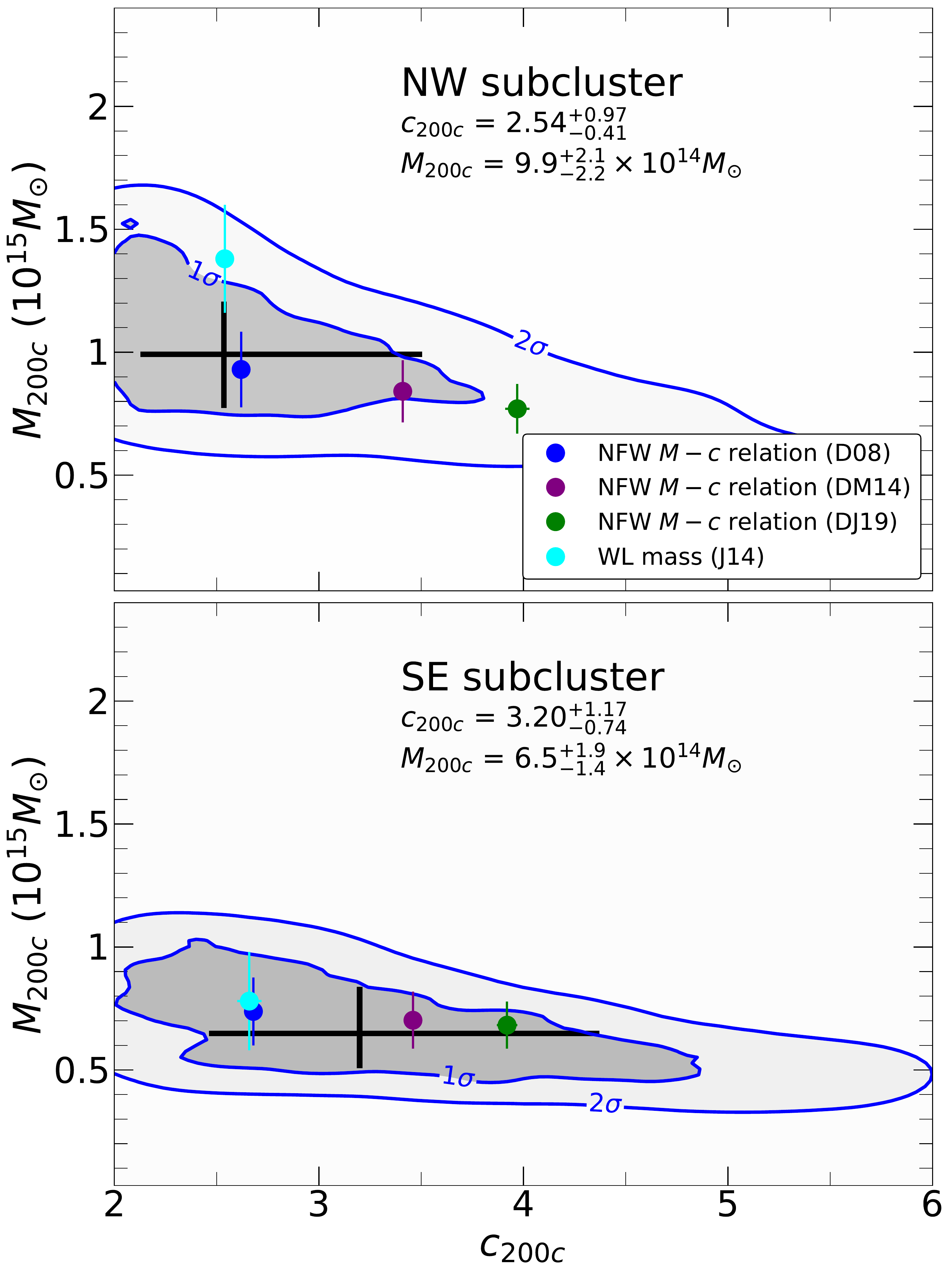}
\caption{Posterior distribution from 100,000 MCMC samples of our two-parameter mass determination when we assume that the centers are at the two mass peaks. The black crosses show the best-fit values and the uncertainties of $c_{200c}$ and $M_{200c}$ after one-dimensional marginalization. 
For comparison, we display the results when mass-concentration $M-c$ relations are used and our previous WL result (J14) using D08 $M-c$ relation. 
These fitting results are consistent with the mass estimates obtained from our two-parameter sampling.}
\label{fig:MCMC_figs}
\end{figure}

\subsection{Mass Estimation} \label{mass_estimates}
Given the cosmological importance of \elgordo, its accurate mass estimation is one of the highest priorities in our study. 
Since \elgordo~is comprised of two massive post-collision subclusters, care must be taken to minimize biases in our mass estimation. The following issues are considered. 
First, we model the WL signal as a superposition of two halos. Treating merging clusters as a single halo can induce a severe bias in mass estimation. 
Second, we do not use the mass-concentration relations. Because \elgordo~is very likely a post-collision system, the mass-concentration relations obtained from ensemble averages of clusters from numerical simulations may not be valid in this system. 
Third, we do not fix the centroid while fitting. 
Typically, WL masses are estimated under the assumption that the center is known. Many studies choose BCGs as cluster centers, which may be safer choices than X-ray peaks in merging clusters. However, as numerical simulations (e.g., \citealt{Martel2014}) demonstrate, a significant fraction of merging clusters have their BCGs offset from the cluster halo centers. 
In this study, we allow the center of each subcluster to move within $10$\arcsec~(corresponding to $\mytilde80$ kpc at $z=0.87$) radius from its peak position during fitting. 
This $10$\arcsec~prior interval approximately corresponds
to the 2~$\sigma$ range in the centroid uncertainties and also encloses the centers defined by other (e.g., X-ray, SZ, galaxies, etc.) data (see Figure~\ref{fig:centroids}). 
Fourth, we exclude the WL signal at $r<30\arcsec$ (\mytilde238~kpc) because in the core there exist both theoretical (e.g., profile shape) and observational (e.g., member galaxy contamination) issues.

Inevitably, we must assume a particular halo model in order to convert projected lensing signals into spherical masses. 
We choose the Navarro-Frenk-White (NFW, \citealt{NFW1997}) profile, although as mentioned above, we do not assume any mass-concentration relation. Aperture mass densitometry (AMD, \citealt{Fahlman1994}) provides a parameter-free method to estimate projected masses. 
We compare the projected masses from the AMD and our best-fit NFW models to assess the potential systematics due to the halo profile model assumption.

We use a Bayesian sampling method to determine the mass distribution of \elgordo~based on the following log-likelihood ($\mathcal{L}$):
\begin{equation}
\mathcal{L} = \sum_{i} \sum_{j=1,2} \frac{ [ g^{pre}_j - g^{obs}_j]^2 } {\sigma_{SN}^2 + \sigma_{e,i}^2}, \label{model2D}
\end{equation}
where 
\begin{align*}
g^{pre}_j = &~g^{pre}_j(x_i,y_i,z_s,M_{NW},c_{NW},x_{NW},y_{NW}, \\
~&~~~~~~~M_{SE},c_{SE},x_{SE},y_{SE})
\end{align*}
and $g^{obs}_j = g^{obs}_j(x_i,y_i)$ are the $j^{th}$ components of the predicted and observed reduced shears of the $i^{th}$ galaxy, respectively, at the position $(x_i,y_i)$ and with source redshift $z_s$, for cluster masses $M_{NW}$ and $M_{SE}$ with concentrations $c_{NW}$ and $c_{SE}$ at the cluster centers ($x_{NW}$, $y_{NW}$, $x_{SE}$, and $y_{SE}$). 
The quantities $\sigma_{SN}$ and $\sigma_{e,i}$ are the ellipticity dispersion (shape noise) and the measurement uncertainty. We set $\sigma_{SN}$ to 0.25 whereas $\sigma_{e,i}$ is determined for each galaxy from the shape measurement.

To determine the posteriors of masses and concentrations ($M_{NW}$, $M_{SE}$, $c_{NW}$, and $c_{SE}$), we perform Markov Chain Monte Carlo (MCMC) sampling. We use flat priors for both concentration $c_{200c}$ and mass $M_{200c}$. 
The prior intervals of $c_{200c}$ and $M_{200c}$ are set to $2 < c < 6$ and $10^{13} < M_{200c} < 10^{16}$, respectively. Given that most simulated high-redshift ($z\sim1$) clusters have concentration parameters within the interval (e.g., see Figure 1 of \citealt{Ludlow2014}), these prior intervals are sufficiently large. 
As we discuss in Appendix~\ref{change_priors}, the impact of the prior choice on our mass measurement is not significant, amounting to $\mytilde10$\% in mass, which is only $\mytilde40$\% of the statistical uncertainty.

Figure~\ref{fig:MCMC_figs} displays the posterior distributions obtained from our 100,000 MCMC samples for mass and concentration parameters of the NW and SE components. 
We obtain $M_{200c} = 9.9^{+2.1}_{-2.2}~\times$~\solarm~and $6.5^{+1.9}_{-1.4}~\times$~\solarm~for NW and SE, respectively. The posteriors of the concentration parameters are not well-constrained because the signals are not sensitive to them. 
We summarize the marginalized parameters in Table~\ref{table_2PMD_MCMC}.

For fair comparison with previous studies, we repeat the mass measurement using three $M-c$ relations: \citet[][D08]{Duffy08}, \citet[][DM14]{DM14}, and \citet[][DJ19]{DJ19}. These results are displayed in Figure~\ref{fig:MCMC_figs} and summarized in Table~\ref{table_1PMD}. 
Also shown in Table~\ref{table_1PMD} are the masses when we use the singular isothermal sphere (SIS) model, which of course is not the best assumption for mass estimation, but nevertheless is convenient for the prediction of the velocity dispersion. 
We find that all these masses are consistent with one another and also with the one obtained without the $M-c$ relation. 
The $1\sigma$ uncertainties from the $M-c$ relations are smaller than those obtained without any $M-c$ relation because of reduction in the number of free parameters. We refer readers to Appendix~\ref{check_2Dmodels} for detailed comparison.

Following J14, we obtain the total mass of \elgordo~by summing the contributions from the two (NW and SE) components while assuming that the two halos are at equal distance from us; the total mass is not sensitive to the geometry for moderate viewing angles (e.g., $\mytilde10$\% mass decrease at $\alpha \sim 60$\degr; see Figure 15 of J14). 
For each set of the two NFW parameters, we compute the center of mass and determine the radius of a sphere, which encloses $M_{200}$. Readers are referred to Appendix~\ref{mass_total} and J14 for details. 
The total mass is estimated to be $M_{200c} = 2.13^{+0.25}_{-0.23}~\times$~\solarmA~($M_{200a} = 2.37^{+0.29}_{-0.27}~\times$~\solarmA).

The mass estimates presented here are based on the assumption that 1) each halo follows a spherical NFW model and 2) no large scale structures affect our measurements. As discussed in J14, studies with numerical simulations showed that these systematics can contribute additional $\mytilde$20\% uncertainty for massive halos such as \elgordo. Since this systematic error is not correlated with the statistical error quoted above, the total error budget should be estimated by adding the two values in quadrature.

\begin{table*}
\begin{center}
\caption{Mass estimates of \elgordo~without mass-concentration relation$^a$\label{table_2PMD_MCMC}}
\begin{tabular}{cccccccc}
\tableline
\tableline
 & \multicolumn{2}{c}{Centroid} \\ 
 \colhead{Component} & \colhead{R.A.} & \colhead{DEC.} & \colhead{$c_{200c}$} & \colhead{$R_{200c}$} & \colhead{$M_{200c}$} & \colhead{$R_{500c}$} & \colhead{$M_{500c}$} \\ 
 & \colhead{(J2000)} & \colhead{(J2000)} & & \colhead{(Mpc)} & \colhead{($10^{14}~M_{\sun}$)} & \colhead{(Mpc)} & \colhead{($10^{14}~M_{\sun}$)} \\ [1.2ex]
\hline
\\[0.1ex]
NW & 01:02:51.23 & -49:15:2.56 & $2.54^{+0.97}_{-0.41^{b}}$ & $1.5\pm0.1$ & $9.9^{+2.1}_{-2.2}$ & $0.9\pm0.1$ & $6.2\pm1.2$ \\ [1.2ex]
SE & 01:02:56.95 & -49:16:21.86 & $3.20^{+1.17}_{-0.74^{b}}$ & $1.3\pm0.1$ & $6.5^{+1.9}_{-1.4}$ & $0.8\pm0.1$ & $4.3^{+1.0}_{-0.9}$ \\ [1.2ex]
Full system & 01:02:53.49$^c$ & -49:15:33.96$^c$ & ... & $2.0\pm0.1$ & $21.3^{+2.5}_{-2.3}$ & $1.3\pm0.1$ & $14.2^{+1.5}_{-1.4}$ \\ [1.2ex]
\hline
\hline
\tableline
\end{tabular}
\end{center}
\tablecomments{a. We obtain mass estimates and concentration parameters by fitting the two-dimensional models to individual galaxy shapes via a Markov Chain Monte Carlo analysis (see text).\\
b. The lower bound of concentration parameter is set to be 2 (see Figure~\ref{fig:MCMC_figs}). \\
c. The center of mass for the two subclusters}
\end{table*}

\begin{table*}
\begin{center}
\caption{Mass estimates of \elgordo~using the NFW profile with various mass-concentration relations and the SIS profile \label{table_1PMD} }
\begin{tabular}{cccccccccc}
\tableline
\tableline
      & \multicolumn{6}{c}{2D NFW$^{a}$} & \multicolumn{3}{c}{2D SIS$^{b}$} \\ 
      \cmidrule(lr){2-7}\cmidrule(lr){8-10}
 \colhead{Component} & \colhead{$R_{200c,D08}$} & \colhead{$M_{200c,D08}$} & \colhead{$R_{200c,DM14}$} & \colhead{$M_{200c,DM14}$} & \colhead{$R_{200c,DJ19}$} & \colhead{$M_{200c,DJ19}$} & \colhead{$\sigma_{v}$} & \colhead{$R_{200c,SIS}$} & \colhead{$M_{200c,SIS}$} \\
 & \colhead{(Mpc)} & \colhead{($10^{14}~M_{\sun}$)} & \colhead{(Mpc)} & \colhead{($10^{14}~M_{\sun}$)} & \colhead{(Mpc)} & \colhead{($10^{14}~M_{\sun}$)} & \colhead{(\kms)} & \colhead{(Mpc)} & \colhead{($10^{14}~M_{\sun}$)} \\ [1.2ex]
\hline \\[0.1ex]
NW & $1.5\pm0.1$& $9.3^{+1.7}_{-1.4}$& $1.4\pm0.1$& $8.4^{+1.4}_{-1.2}$& $1.4\pm0.1$& $7.7\pm1.0$ & $1061\pm47$ & $1.3\pm0.1$ & $7.0\pm0.9$ \\ [1.2ex]
SE & $1.4\pm0.1$& $7.4^{+1.5}_{-1.2}$& $1.3\pm0.1$& $7.0^{+1.3}_{-1.1}$& $1.3\pm0.1$& $6.8\pm1.0$ & $1017\pm48$ & $1.3\pm0.1$ & $6.2^{+0.9}_{-0.8}$ \\ [1.2ex]
Full system & $2.0\pm0.1$ & $22.6^{+4.5}_{-4.0}$ & $1.9\pm0.1$ & $19.5^{+3.6}_{-3.0}$ & $1.8\pm0.1$ & $18.1^{+2.3}_{-2.5}$ & ... & ... & ... \\ [1.2ex]
\hline
\hline
\tableline
\end{tabular}
\end{center}
\tablecomments{a. We apply three mass-concentration relations: \citealt{Duffy08} (D08), \citealt{DM14} (DM14), and \citealt{DJ19} (DJ19) \\
b. We do not use the SIS model when determining total mass.}
\end{table*}

\section{Discussion} \label{section_discussion}
\subsection{Mass Comparison with Previous Studies}
\elgordo~is the subject of a number of studies and here we compare our mass estimate with previous results. 
In general, gravitational lensing is considered to be a superior method over others because it does not rely on the dynamical status of the system. 
However, one should not ignore the fact that lensing-based mass estimation is subject to different systematics. 
We first discuss the comparison with the results from non-lensing studies (\textsection\ref{mass_comparison_non_lensing}) and then with the strong-lensing (\textsection\ref{mass_compare_SL}) and weak-lensing (\textsection\ref{mass_compare_WL}) results.

\begin{figure}
\centering
\includegraphics[width=85mm]{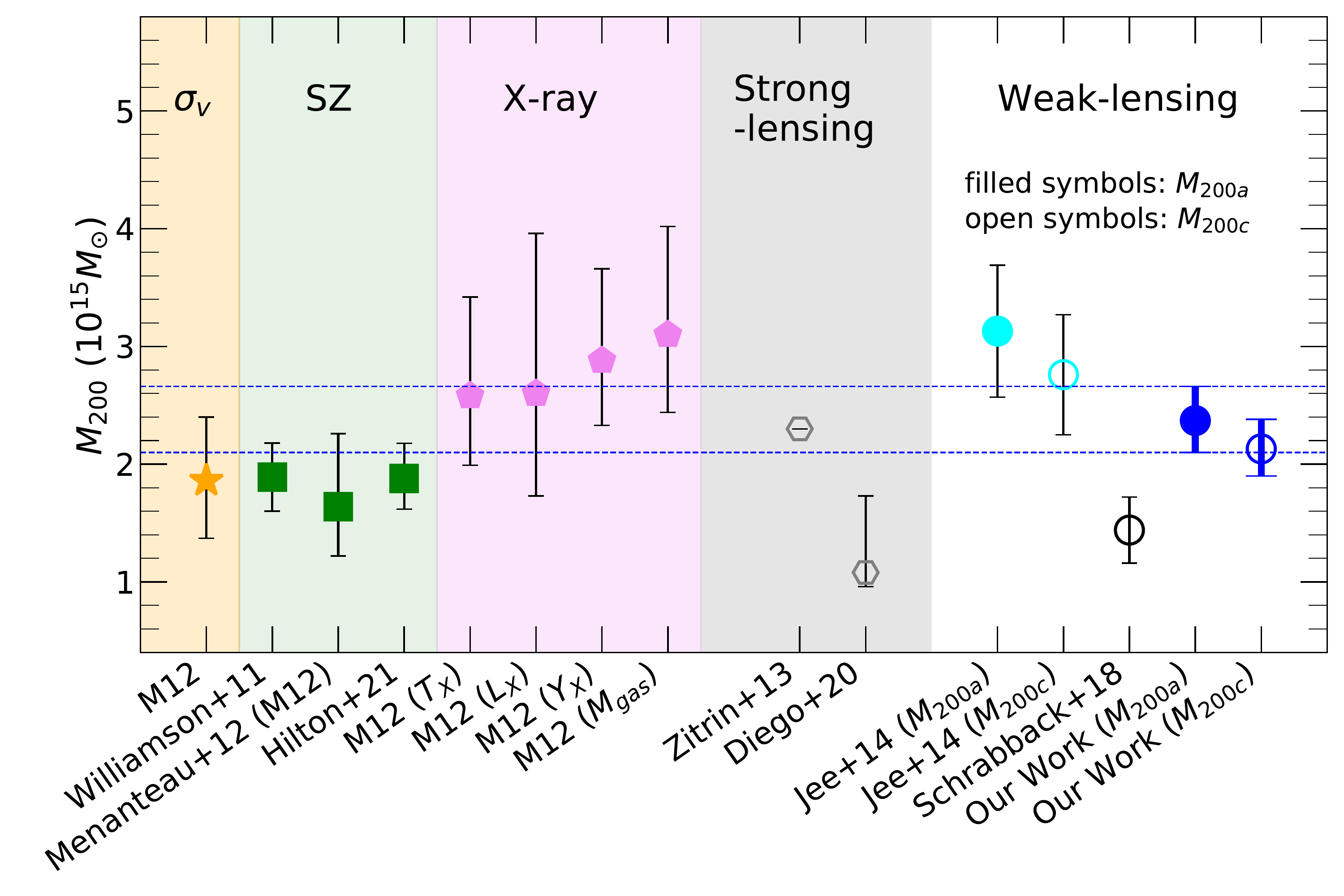}
\caption{Comparison of mass estimates of \elgordo~from various proxies. 
We adopt the best-fit 2D NFW results without $M-c$ relation as our fiducial mass estimates (blue circles). Our WL mass estimates ($M_{200c}$ and $M_{200a}$) tend to be located in the midpoint of the previous lensing results as well as the proxies that rely on the assumed properties of the gas and galaxies (see text for detailed discussion). We only indicate the $1\sigma$ statistical uncertainties of all measurements.}
\label{fig:mass_compare}
\end{figure}

\subsubsection{Previous Non-lensing Studies} \label{mass_comparison_non_lensing}
The ACT SZ discovery of \elgordo~(\citealt{Menanteau2010}; \citealt{Marriage2011}) also provided evidence that the cluster was exceptionally massive: $M_{200} > 10^{15}\ M_\odot$ based on an X-ray luminosity estimate. 
Later \cite{Williamson2011} measured the virial mass of $M_{200a} = (1.89\pm0.29)~\times~$\solarmA~based on the South Pole Telescope (SPT) data. A similar ($1.64^{+0.62}_{-0.42}~\times$~\solarmA) mass is reported by M12 with the ACT data. 
Recently, \cite{Hilton2021} quoted a mass of $M_{200a} = 1.88^{+0.30}_{-0.25}~\times$~\solarmA~from their SZ analysis, which makes \elgordo~the most massive cluster at $z > 0.5$ in their $13211$ sq. deg survey. 
These SZ measurements used the scaling relation between $yT_{CMB}$ and $M_{200a}$, which in turn depends on a calibration with previous X-ray mass estimates.

There exist multiple ways to obtain cluster masses with X-ray data. 
Using the gas mass ($M_{gas}$), product of gas mass and temperature ($Y_X$), luminosity ($L_X$), and temperature ($T_X$), M12 reported X-ray masses ranging from $\mytilde2.58~\times$~\solarmA~to $\mytilde3.10~\times$~\solarmA.

Velocity dispersion based mass estimates are also provided by M12 from 89 spectroscopic members. 
Using the $M-\sigma_{DM}$ relation of \cite{Evrard2008}, they obtained $M_{200a}= 1.86^{+0.54}_{-0.49}~\times$~\solarmA~by treating \elgordo~as a single halo. 
They determined the individual masses of the NW and SE clusters to be $M_{200a}=1.76^{+0.62}_{-0.58}~\times$~\solarmA~and $1.06^{+0.64}_{-0.59}~\times$~\solarmA, respectively.

Figure~\ref{fig:mass_compare} summarizes the above results in methodological and chronological order. 
Compared with the current new WL mass, they are statistically consistent. 
However, since the consistency here is mainly due to the large statistical errors, one should not interpret the consistency as justifying the use of mass proxies under the assumption of a single halo. 
Because \elgordo~is comprised of two systems with dynamically disrupted structures, treating the system as a single halo is an important source of systematics in mass estimation. 
Although mass proxies are typically calibrated against theory or simulations (e.g., the velocity dispersion from \citealt{Evrard2008}), exceptional mergers like \elgordo~are rarely found in the simulations or as significant outliers in the scaling relations.

\subsubsection{Previous Strong-lensing Studies} \label{mass_compare_SL}
The first SL study was presented by \cite{Zitrin2013}. 
From their light-traces-mass approach, they analyzed the SL features within a $\mytilde500$ kpc radius of each subcluster using the early (PROP ID 12755) \HST/ACS imaging data and reported an extrapolated mass of $M_{200c}\sim2.3~\times$~\solarmA. 
\cite{Diego2020} obtained a factor of two lower mass ($M_{200c} = 1.08^{+0.65}_{-0.12}~\times$~\solarmA) from their free-form lens model using ten ACS and WFC3/IR filters (PROP ID 12477, ID 12755, and ID 14096).

The limitations of these SL mass estimations include the large extrapolation to $R_{200}$ and a lack of secure redshift information of the strongly-lensed high-redshift galaxies. 
\cite{Cerny2018} claimed that the use of photo-$z$ leads to a systematic bias of $\mytilde15$\% in mass. Currently, none of the SL features in \elgordo~has a spectroscopic redshift, which makes the mass measurement itself within the small area uncertain and also extrapolation to the virial radius $\mytilde2$~Mpc inaccurate. 
The factor of two discrepancy between \cite{Zitrin2013} and \cite{Diego2020} might illustrate the difficulty in using the SL data to estimate the virial mass.

\begin{figure}
\centering
\includegraphics[width=85mm]{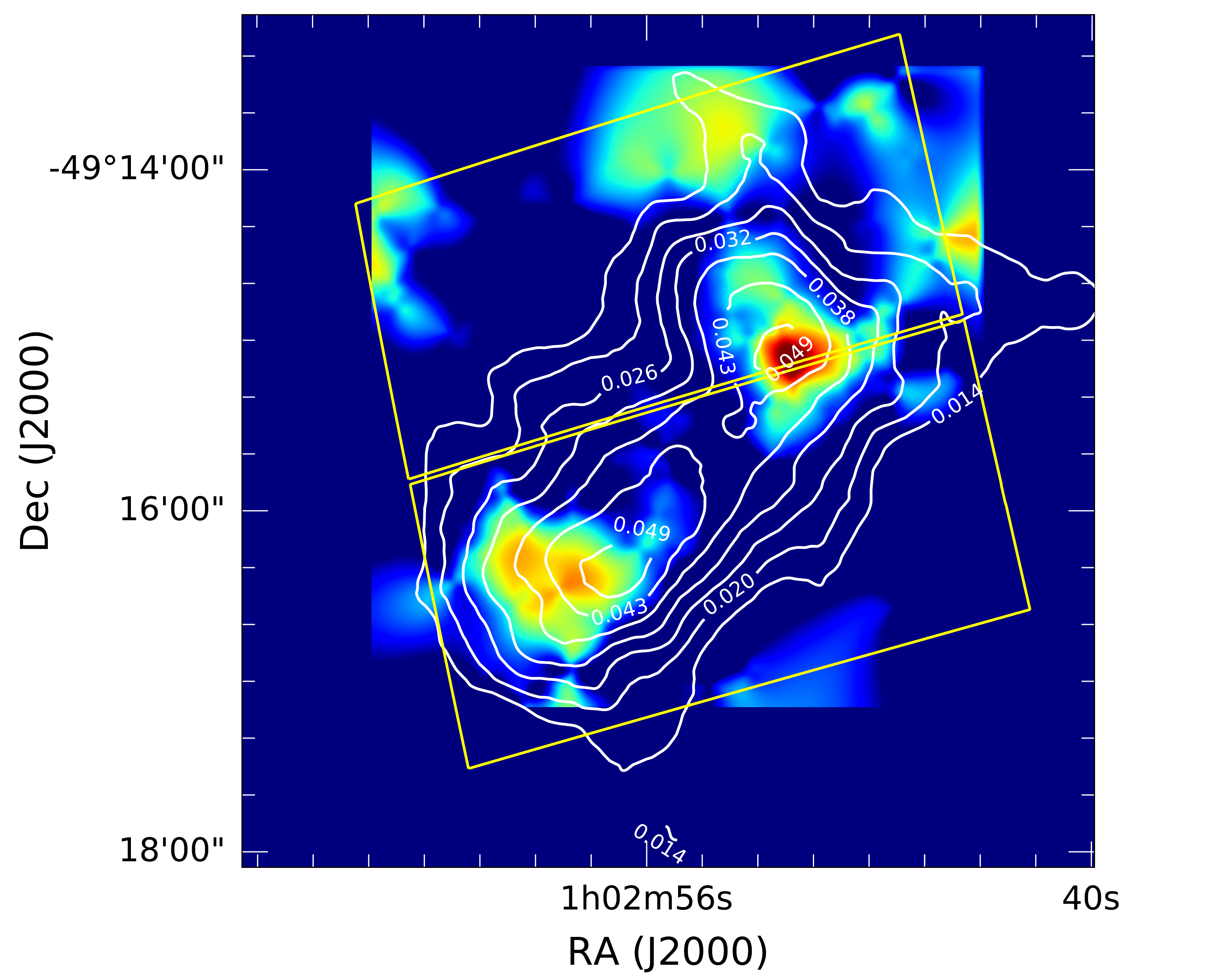}
\caption{F814W-only Mass reconstruction. 
Color-coded is the surface mass density ($\kappa$) obtained from the ACS/F814W data alone. 
The footprint of the ACS/F814W pointing is indicated with yellow solid line. 
White contours show our fiducial mass map obtained with ten ACS and WFC3/IR filters. 
Although the source number density becomes much smaller ($\mytilde42$~\persqarcmin) with the use of the ACS/F814W data alone, both the NW and SE components are clearly detected. 
Note that this result differs from that of \cite{Harvey2015}.}
\label{fig:mass_map_F814W}
\end{figure}

\begin{figure}
\centering
\includegraphics[width=85mm]{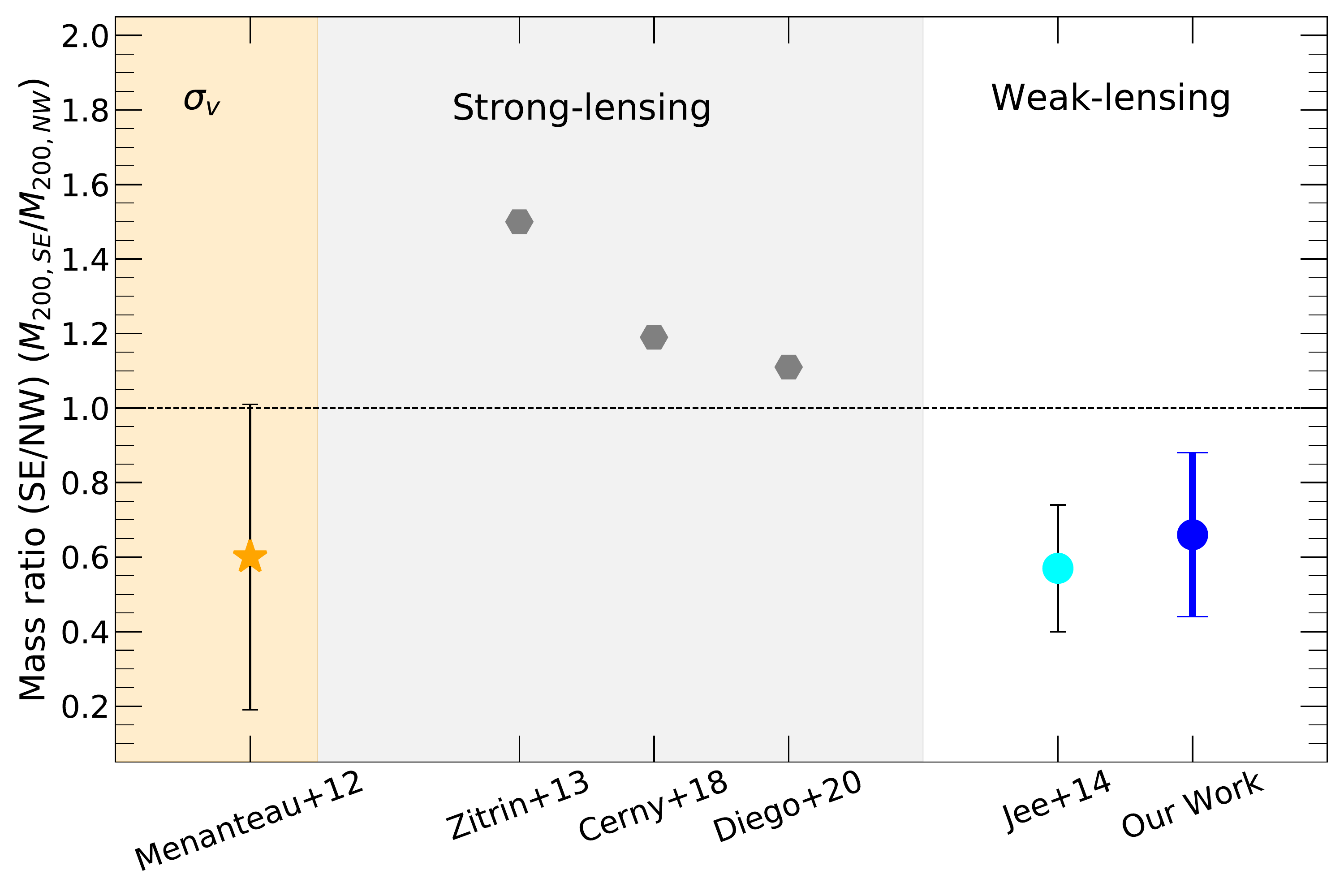}
\caption{Comparison of the subcluster mass ratio measurements ($M_{200}^{SE}/M_{200}^{NW}$).
While SL results (gray hexagons) show that the SE component is more massive, velocity dispersion (orange star) and WL (cyan and blue circles) support the opposite case. 
We argue that the discrepancy might be due to the SE subcluster mass distribution being more compact (see text for details). 
Note that the SL data points do not have error bars because the papers do not quote the uncertainties of their masses.}
\label{fig:mass_ratio_compare}
\end{figure}

\subsubsection{Previous Weak-lensing Studies} \label{mass_compare_WL}
J14 provided the first WL analysis of \elgordo~with the \HST/ACS imaging data covering the central 45~\sqarcmin~area. 
As discussed in \textsection\ref{section_results}, the mass map of J14 is in good agreement with our result. Here we compare only the masses. 
The central value of our virial mass ($M_{200c} = 2.13^{+0.25}_{-0.23}~\times$~\solarmA) for the entire system is $\mytilde23$\% lower than that of J14; when individual masses are concerned, our NW (SE) mass is $\mytilde28$\% ($\mytilde17$\%) lower. 
Although the $1\sigma$ error bars between the two studies overlap, the differences are non-negligible considering that the two studies share a significant number of common background galaxies (the shape noises are highly correlated). One potential source of the difference is the $M-c$ relation employed by J14. We observe that using the relation results in the mass increasing by $\mytilde6$\%, which alone cannot fully explain the $\mytilde23$\% difference. 
Another possible candidate is the inclusion of WFC3/IR shapes. As we demonstrated in \S\ref{section_PSF_shape}, the difference in shape between ACS and WFC3/IR is less than $3$\%, which shows that including WFC3/IR sources cannot be a cause of the reduction in mass estimation. 
Since the two studies rely on the same control fields to determine the redshift of the source population and use the same shear calibration procedure, we argue that the mass difference is mainly caused by the increase of the field size by a factor of 2.7, which enables us to estimate the total mass without any extrapolation.

\cite{Harvey2015} performed WL analysis of \elgordo~with the ACS/F814W data alone. Their mass reconstruction did not detect any mass peak around the NW component (see Figure 2 of \citealt{Harvey2015}). Since the association of the NW component with a large mass concentration is clear in other studies (e.g., \citealt{Zitrin2013}; \citealt{Diego2020}; J14; \citealt{Schrabback2018}), this absence is surprising. 
As \cite{Wittman2018} noted, while they found two mass peaks, their northeast mass peak coincides with the chip gap between the ACS detectors and the suggested merger axis conflicts with the position of radio relics discovered by \cite{Lindner2014}. 
When we repeated our WL analysis with the ACS/F814W data alone, we still detected both NW and SE mass peaks, which closely resembles the current mass map (Figure~\ref{fig:mass_map_F814W}).

The mass reconstruction of \cite{Schrabback2018} is similar to our result, detecting the bimodality. 
However, they assumed a single halo in deriving the total masses using the X-ray and SZ peaks as the cluster centers. They quote $M_{200c} = 1.11^{+0.29}_{-0.28}~\times$~\solarmA~($1.44^{+0.28}_{-0.28}~\times$~\solarmA) for the choice of the X-ray (SZ) peak. 
As mentioned in \textsection\ref{mass_estimates}, assuming a single halo for a binary cluster can be a significant source of systematics in the total mass estimation. 
Another notable difference is the much lower source density ($\mytilde20$~\persqarcmin) in \cite{Schrabback2018}, who applied a tighter color cut in source selection. \\

\subsubsection{Which Mass Clump is Heavier?} \label{mass_ratio_inconsistency}
The mass ratio between the two merging components in a binary collision is one of the key quantities needed for reliable reconstruction of the merging scenario. 
In general, the less massive component experiences a greater relative velocity with respect to the center of mass, and in X-ray the structure appears as a bullet-like feature. In terms of the X-ray surface brightness, the less massive component, which often has a cooler (thus denser) core, is often brighter (e.g., \citealt{Clowe2006}; \citealt{Jee2016}; \citealt{Golovich2017}). 
Also, the greater Mach number of induced merger shocks due to a higher collision velocity with respect to the center of mass can lead to a more prominent radio relic (see Figure 14 of \citealt{vanWeeren2019}).

Interestingly, \elgordo~is a binary merging cluster, which does not have a consensus on which component is more massive. We compare the mass ratio measurements from various studies in Figure~\ref{fig:mass_ratio_compare}. 
The three SL studies assert that the SE component is more massive whereas the three WL studies including the current one and the dynamical study favors the opposite case; although \cite{Schrabback2018} do not present individual masses, their mass map indicates that the surface mass density in the NW region is significantly higher.

We suspect that the difference in mass ratio values between strong and weak lensing measurements might originate from the use of the extrapolation in SL studies because the projected mass density of the SE component at small radii (within the SL regime $\mytilde500$~kpc) is estimated to be higher. 
However, without secure redshifts of the SL features, currently it is difficult to dispute or support the previous SL results.

With our WL data we performed an experiment to test whether or not the WL signal around the SE cluster is indeed higher at small radii. 
We cropped our WL source catalog in such a way that only sources at $r<1.5$~Mpc from each mass peak remain with the resulting number of sources being $2081$. 
Our simultaneous two-NFW-profile fitting to this cropped catalog finds that the SE component is $1.5\pm0.6$ times more massive than the NW one. 
This experiment suggests that in the central region the projected mass of the SE cluster might be higher as indicated by the SL studies, although the significance is somewhat marginal.

The X-ray morphology of \elgordo~provides circumstantial evidence that the NW component is more massive. 
Numerical simulations (e.g., \citealt{Mastropietro2008}; \citealt{Molnar2015}; \citealt{Moura2021}) have shown that after a major merger core passage (i.e., small impact parameter) the dark matter and ICM associated with the more massive (thus hotter in X-ray) component tend to disperse more severely while the ones with the less massive component (thus colder in X-ray) maintain their compactness; the dispersion is stronger in gas than in dark matter. 
The current {\it Chandra} image of \elgordo~reveals only a single peak at the position of the SE component (Figure~\ref{fig:mass_map_HST}). 
The absence of a corresponding X-ray peak near the NW mass/galaxy clump may be due to the aforementioned dispersion after the core passage. 
The temperature map structure is also consistent with this scenario. The X-ray peak region is relatively cold. Perhaps, the low mass density in the central region of the NW mass clump, if true, may be caused by the post-merger dispersion.

The dynamical study of the cluster member galaxies based on optical/IR observations supports our WL result. 
M12 investigated the spatial distribution of the cluster member galaxies selected by spectroscopic and/or photometric redshift. 
They found that more cluster member galaxies (and thus more stellar mass density) are associated with the NW component (see Figure 8 of M12). They also measured the dynamical mass of each component based on the velocity dispersions of cluster member galaxies and the $M-\sigma_{DM}$ relation of \cite{Evrard2008}, which suggests that the NW component is more massive (Figure~\ref{fig:mass_ratio_compare}).

\begin{figure}
\centering
\includegraphics[width=85mm]{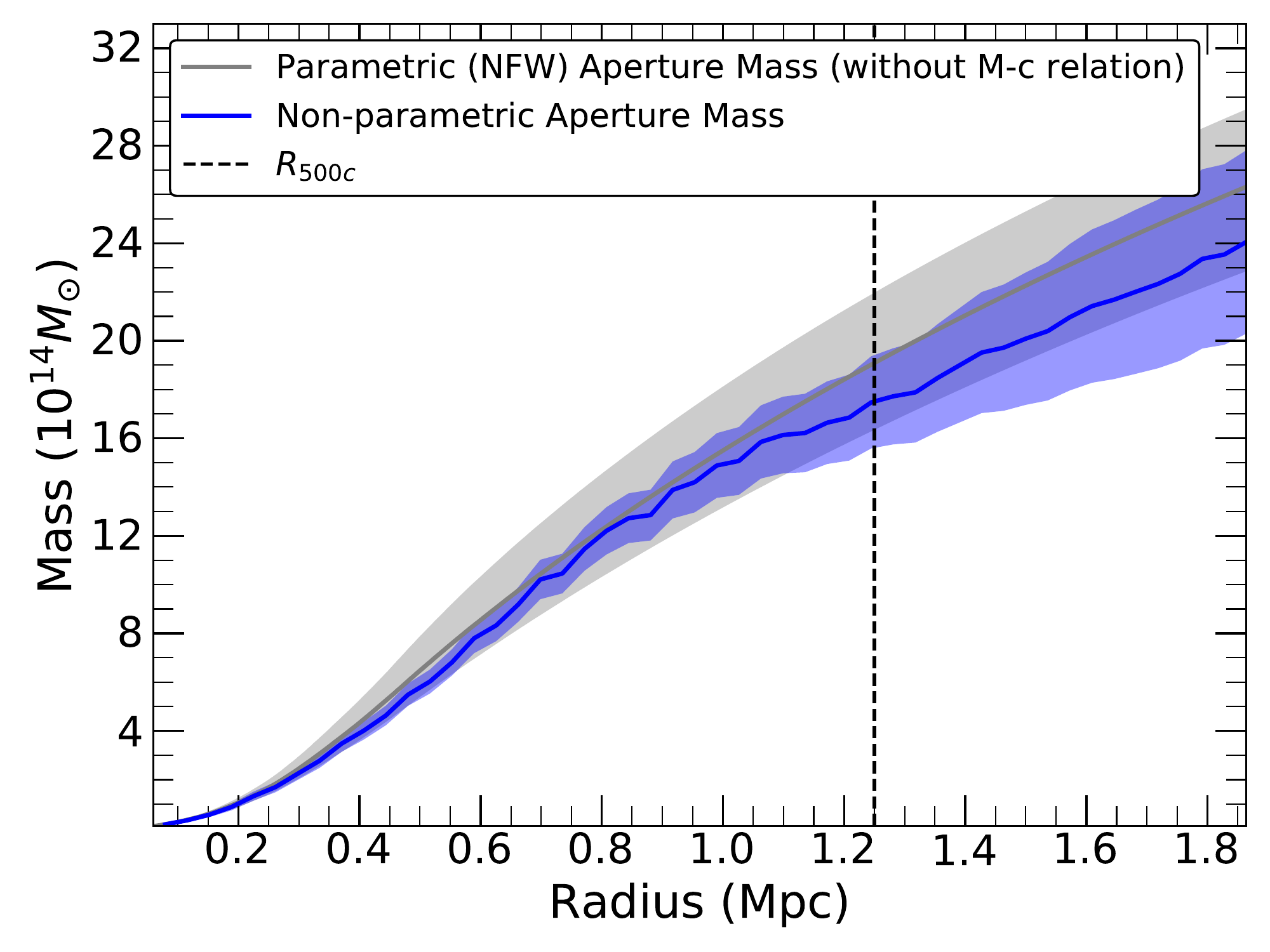}
\caption{Projected mass profile of \elgordo. 
The non-parametric (parametric) result is shown in blue (gray). Shaded regions depict $1\sigma$ uncertainty intervals. The comparison shows that both results are highly consistent, which indicates that our NFW assumption does not give rise to a significant model bias in cluster mass estimation.}
\label{fig:AMD}
\end{figure}

\subsection{NFW Model Bias} \label{systematics}
In \S\ref{mass_estimates}, we determine the mass of \elgordo~by simultaneously fitting two NFW models. 
We perform the mass estimation without any $M-c$ relation because the merger may cause the density profile to deviate from the relation. Although the procedure allows us to marginalize over a wider range of possibilities, the model freedom is still limited to the NFW paradigm. \cite{Becker2011} claimed that the intrinsic scatter in mass measurement from the NFW assumption is $\mytilde20$\% for massive clusters. 
Here we present the cluster masses from AMD and compare the results with the NFW predictions. Since AMD provides a nearly model-independent projected mass, any significant difference between the two can be used as a useful diagnostic of the NFW model bias.

AMD measures the projected mean overdensity $\zeta_c$ in units of the critical surface density $\Sigma_{c}$ (Equation~\ref{eqn_sigma_c}) within a given circular aperture at $r=r_1$ relative to the control annulus defined by $r=r_2$ and $r_{max}$:
\begin{eqnarray}
\zeta_c (r_1, r_2,r_{max}) =  \bar{\kappa}( r \leq r_1) -
\bar{\kappa}( r_2 < r \leq r_{max}) \nonumber \\ = 2 \int_{r_1}^{r_2} \frac{
 \left < \gamma_T \right > }{r}dr  + \frac{2}{1-r_2^2/r_{max}^2}
\int_{r_2}^{r_{max}} \frac{ \left <\gamma_T \right >}{r} dr, \label{eqn_AMD}
\end{eqnarray}
\noindent
where $\left < \gamma_T \right>$ is the azimuthally-averaged tangential shear. 
The tangential shear $g_T$ is computed by
\begin{equation}
g_T = -e_1 \cos 2\theta - e_2 \sin 2\theta \label{tan_shear},
\end{equation}
\noindent
where $e_{1(2)}$ is the real (imaginary) part of the ellipticity obtained from Equation~\ref{e1e2} and $\theta$ is the position angle of each source galaxy with respect to the center. 
Because $\gamma$ is obtained only after $\kappa$ is known [i.e., $g=\gamma/(1-\kappa)$], we begin with $\left < g_T \right >$ by assuming $\kappa=0$ and then replace $\left < g_T \right >$ with $\left < \gamma_T \right>$ using the estimate on $\kappa$.
The procedure converges quickly.
We set the inner and outer radii of the control annulus to be $240$\arcsec~($\mytilde1.9$ Mpc) and $260$\arcsec~($\mytilde2.1$ Mpc), respectively, which is completely covered by our new wide-field ACS observations. 
The average surface density within the control annulus is estimated to be $\bar{\kappa}=0.017$ from our best-fit model in \S\ref{mass_estimates}.
We refer readers to \cite{Clowe2000} and \cite{Jee2005} for detail.

Figure~\ref{fig:AMD} shows the cumulative projected mass profile from this AMD approach. The center of the profile is chosen to be the projected center of mass of the two merging components. The shaded region shows the $1\sigma$ uncertainties from bootstrapping. 
For comparison, we project the two (NW and SE) best-fit NFW profiles (Table~\ref{table_2PMD_MCMC}) and display the sum. 
Since the two results are highly consistent within the range ($r\lesssim1.9$~Mpc) probed by AMD, we do not find any evidence for significant model bias caused by our NFW assumption.

\begin{figure}
\centering
\includegraphics[width=85mm]{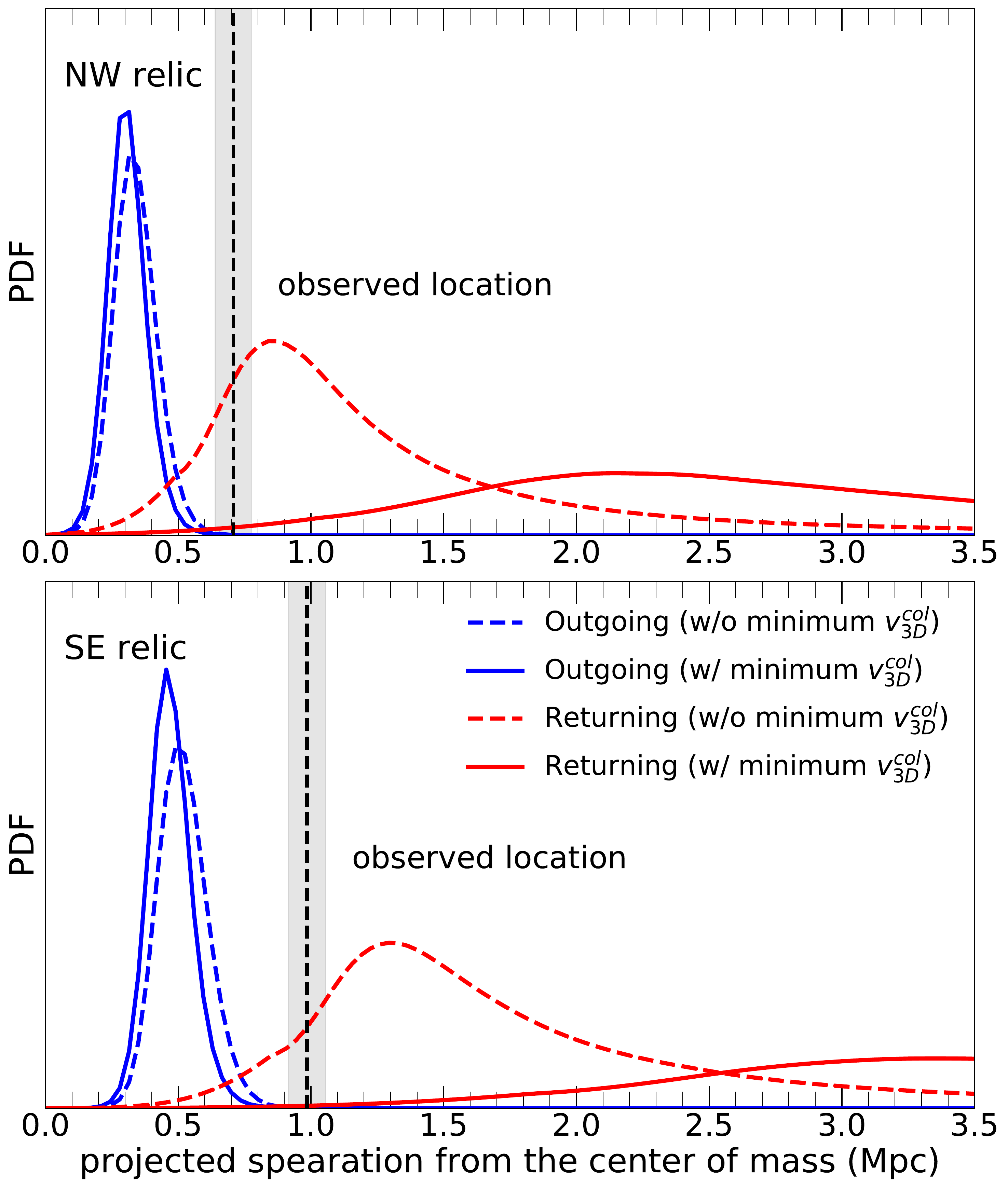}
\caption{Probability density functions (PDFs) of the projected distance of radio relics (top: NW, bottom: SE) from the center of mass. 
When the minimum collision velocity requirement is not imposed, our Monte Carlo simulation results favor the returning scenario (dashed line), which is consistent with the \cite{Ng2015} result. 
Applying the minimum collision velocity requirement produces very different results as shown with solid lines. Neither the outgoing nor returning phase is favored in this case. We argue that the PDFs would be significantly modified if dynamical friction were included (see text). }
\label{fig:merger_scenario}
\end{figure}

\subsection{Merger Scenario} \label{merger_scenario}
The current observation of \elgordo~provides only a snapshot in the long history of its merger. 
Critical parameters in determination of the merger phase and geometry include the X-ray gas morphology, halo masses, and positions and orientations of the radio relics. Here we briefly review the previous efforts to constrain the merger scenario of \elgordo, discuss their strengths and weaknesses, and present our Monte Carlo analysis result.

The hydrodynamic simulations for \elgordo~to date (e.g., \citealt{Donnert2014}; \citealt{Molnar2015}; \citealt{Zhang2015}) mainly focus on reproduction of the morphology of the hot gas seen in X-ray, which is characterized by two prominent cometary tails. These simulations considered only an outgoing phase (i.e., two subclusters moving away from each other after first pericenter crossing). 
Although some of these numerical studies are partly successful in creating the observed X-ray morphology, their infall velocities exceed the theoretical maximum velocity, which is set by a timing argument, as noted by \cite{Ng2015}. 
For example, one setup of the \cite{Zhang2015} simulations starts with an infall velocity of $2500$~\kms~at a separation of $5.5$ Mpc, which results in a collision velocity of $\mytilde4600$~\kms~larger than the (timing argument-based) escape velocity ($\mytilde4200$~\kms). 
Since the two subclusters of \elgordo~must start their freefall toward each other with much smaller masses than the current values, the collision velocity cannot exceed the escape velocity computed with the present masses. 
Therefore, it is not surprising that cosmological studies fail to find any merger analog for \elgordo~(e.g., \citealt{Asencio2021}). 
None of the previous hydrodynamic simulation studies attempt to reproduce the radio relics of \elgordo.

Based on the Monte Carlo Merging Analysis Code \citep[MCMAC;][]{MCMAC}, \cite{Ng2015} estimated the merger parameters of \elgordo~using the radio relic observations, which provide constraints on the viewing angle and collision velocity of the merger. 
The MCMAC takes observed priors on each subcluster's properties (e.g., mass, radial velocity, projected separation, etc.), draws random samples from these priors, and uses them in an analytic model to obtain posteriors of merger parameters. 
Although this approach lacks details compared to time-consuming $N$-body simulations, it enables us to quickly search the merger parameter space for the most probable cases. 
\cite{Ng2015} concluded that the MCMAC analysis strongly favors a returning phase with a collision velocity of $\mytilde2400$~\kms. The observed offset between X-ray and SE mass peak may support this scenario.

\cite{Ng2015} carefully excluded samples with unphysically large collision velocities mentioned above (i.e., exceeding the escape velocity). 
However, they did not exclude unphysically small collision velocities. For example, if the two subclusters start to freefall toward each other at the ``touch" separation (equal to the sum of their virial radii, $r^{NW}_{200c}+r^{SE}_{200c}\sim2.8$~Mpc), the collision velocity at the time of impact exceeds $\mytilde3000$~\kms. 
However, \cite{Ng2015} included samples whose collision velocities are as low as $\mytilde1900$~\kms. This is because they allowed the case even when the two clusters start their freefall at the current projected separation $\mytilde0.7$~Mpc. 
We argue that this case is also unphysical because at a separation of $\mytilde0.7$~Mpc one subcluster's virial radius already encloses the center of the other subcluster. Note that the best-fit collision velocity ($\mytilde2400$~\kms) of \cite{Ng2015} is smaller than the above minimum velocity $\mytilde3000$~\kms.

Another drawback of the \cite{Ng2015} study is the incorrect random number generation for the merger viewing angle $\alpha$. 
In order to make the probability density function (PDF) follow $p(\alpha)\propto\cos \alpha$, the correct transformation should be $\sin^{-1} x$, where $x$ is a random number providing a uniform distribution between 0 and 1. 
In the original version of MCMAC used by \cite{Ng2015}, the transformation was $\propto 1-\cos \alpha$, where $\alpha$ is uniformly distributed between $0$ and $\pi/2$.

We ran MCMAC with the aforementioned two critical updates: exclusion of unphysically small collision velocities and correct viewing angle sampling. 
Figure~\ref{fig:merger_scenario} shows the resulting PDF of the radio relics of \elgordo. Without the minimum velocity constraint (as in the original version), we are able to reproduce the result of \cite{Ng2015}, favoring the returning scenario. 
However, when the minimum collision velocity condition is imposed, the current radio relic locations support neither the outgoing nor the returning case. 
The minimum collision velocity requirement results in higher collision velocities and higher shock propagation velocities. 
For the outgoing phase, this reduces the time-since-collision on average. Thus, the expected radio relic position is only slightly affected because the increase in the shock propagation velocity is nearly offset by the reduction in time-since-collision. 
On the other hand, for the returning phase, the larger collision velocity results in a longer time-since-collision because the subcluster must reach a more distant apocenter before returning. 
Consequently, the radio relic PDF shifts significantly toward larger values.

Our reanalysis of \elgordo~with MCMAC provides results different from \cite{Ng2015}. 
At face value, Figure~\ref{fig:merger_scenario} presents a puzzle for explaining the current position of the radio relics. 
However, we believe that the issue can be resolved when we consider dynamical friction, which is not included in the current analysis. 
Based on comparing the MCMAC and numerical simulation results for the Bullet cluster with a $\mytilde10:1$ mass ratio, \cite{MCMAC} argues that the effect is not significant. However, we find that the effect is non-negligible for \elgordo, where both subclusters are extremely massive. 
In order to assess the effects of dynamical friction, we ran a hydrodynamic simulation of \elgordo~with Gadget v2.0 assuming the best-fit masses of our WL study. 
When the two subclusters freefall from the contact separation, the collision velocity becomes $\mytilde86$\% of the analytic solution provided by MCMAC (i.e., purely based on the potential energy). 
And, more importantly, the time-since-collision value is reduced by a factor of 3-4 for the returning case compared to the MCMAC estimation. 
Therefore, dynamical friction can shift the radio relic PDF for the returning case to the left in such a way that a significant part of the PDF overlaps the observed radio relic position. 
Of course, firm conclusions should await studies with a more complete set of simulations based on various merger configurations. \\

\subsection{Rarity Tests} \label{rarity}
The extraordinary characteristics of \elgordo~such as the extreme mass and large infall velocity at high redshift have stimulated quite a few studies on the compatibility of its existence with the \LCDM~paradigm (e.g., M12; \citealt[][]{Waizmann2012a,Waizmann2012b}; \citealt{Harrison2012}; \citealt{Katz2013}; J14; \citealt{Sahlen2016}; \citealt{Asencio2021}). 
Here, we briefly revisit the issue with our new mass (\S\ref{rarity_mass_only}) and infalling velocity (\S\ref{rarity_mass_velocity}).

\begin{figure}
\centering
\includegraphics[width=85mm]{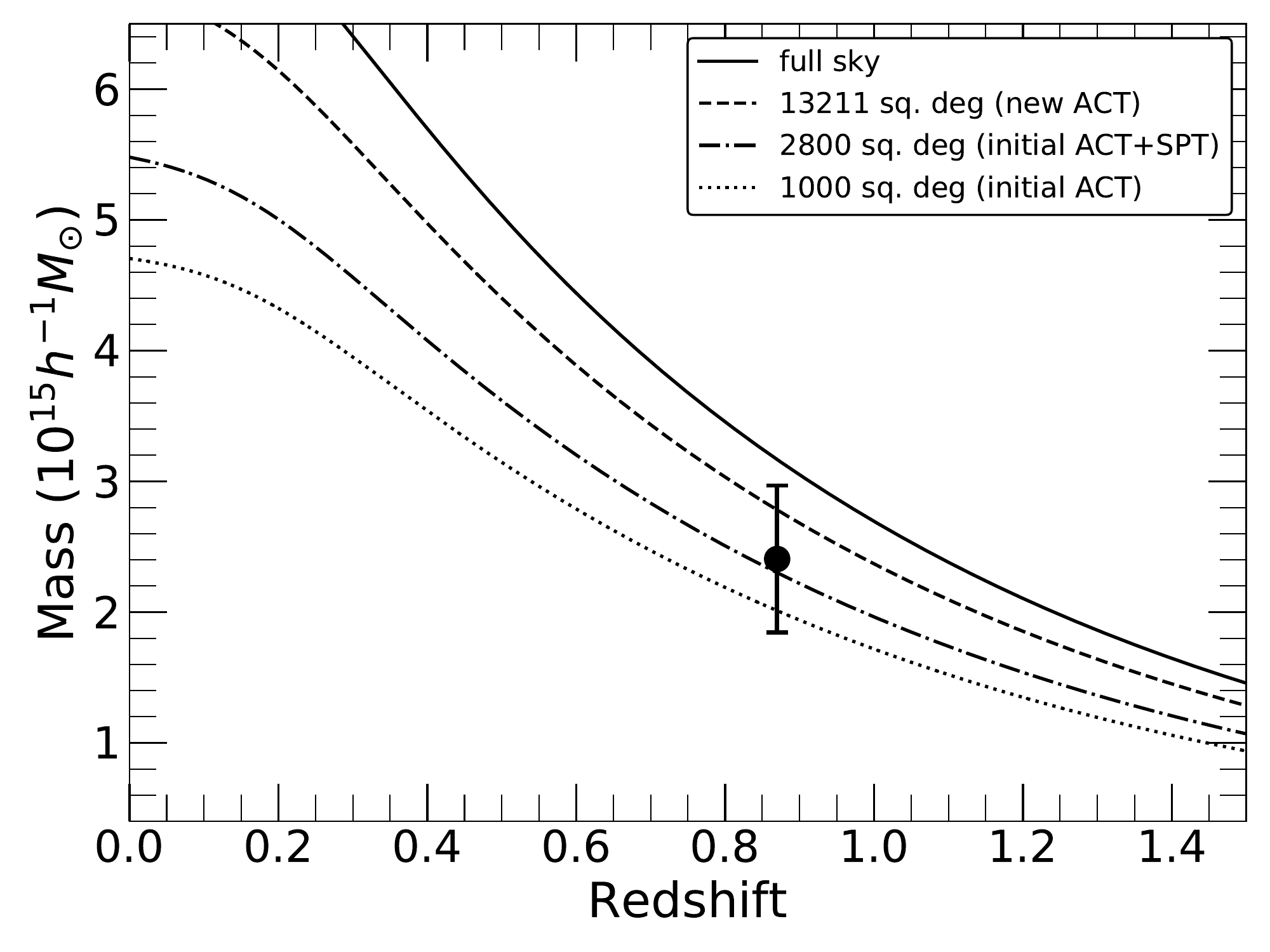}
\caption{Exclusion curves and mass estimates ($M_{200a}$) of \elgordo. 
We display the exclusion curves of $95$\% ($\mytilde2\sigma$) confidence level for both sample and parameter variances for full sky (solid), $13211$ sq. deg (dashed, the current result of ACT survey; \citealt{Hilton2021}), $2800$ sq. deg (dashed-dot, the initial ACT and the SPT surveys; \citealt{Hasselfield2013}; \citealt{Williamson2011}), and $1000$ sq. deg (dotted, the initial ACT survey only; \citealt{Hasselfield2013}).
The filled circle and its error bar are our WL mass estimate and $\mytilde2\sigma$ statistical uncertainty, respectively.}
\label{fig:exclusion_curve}
\end{figure}

\subsubsection{Rarity Test based on the Total Mass} \label{rarity_mass_only}
The conventional measure of a cluster's rarity is the expected number of similarly massive clusters above certain mass and redshift thresholds. 
It is worth mentioning that this approach is one of the least conservative methods and can produce biased results. 
As \cite{Hotchkiss2011} pointed out, the method tends to overestimate the tension. One important cause of the overestimation is that this approach does not include the cases of larger masses at lower redshifts. 
The bias increases further when one also considers the Eddington bias \citep{Eddington1913} due to the extreme steepness of the mass function, and the current uncertainties of the cosmological parameter measurements. 
The expected abundance from the conventional approach is given by 
\begin{equation}
N(M,z) = f_{sky} \int_{z_{min}}^{z_{max}} \frac{dV(z)}{dz} dz \int_{M_{min}}^{M_{max}} \frac{dn}{dM} dM, 
\label{eqn_abundance}
\end{equation}
\noindent
where $f_{sky}$ is the sky coverage fraction\footnote{For example, $f_{sky}\sim0.07$ for a $2800$ sq. deg survey}, $dV/dz$ is the comoving volume element, and $dn/dM$ is the halo mass function. 
In this work, we employ the mass function of \cite{Tinker2008} and the \cite{Planck2016} cosmology. 
We choose the cluster redshift ($z=0.87$) and the $1\sigma$ lower limit of our mass estimate ($M_{200a} = 2.37^{+0.29}_{-0.27}~\times$~\solarmA) for $z_{min}$ and $M_{min}$, respectively. 
With the conservative condition that we assume full-sky coverage and adopt the $1\sigma$ lower limit of the mass estimate, we find that the probability of discovering at least one massive cluster at $M>M_{min}$ and $z>z_{min}$ is $\mytilde9.3$\%. This does not provide serious tension with \LCDM.

\cite{Mortonson2011} proposed an exclusion curve method, which predicts the theoretical maximum mass as a function of redshift at a given cosmology and survey volume. 
The exclusion curve includes both parameter and sample variances of the cosmology within specific confidence levels (CL). 
Figure~\ref{fig:exclusion_curve} shows the exclusion curves for four survey areas ($1000$ sq. deg, $2800$ sq. deg, $13211$ sq. deg, and full-sky) when we set the CL values of the sample variance and parameter uncertainty to 0.95 ($95$\% or $2\sigma$ CL). 
If a cluster lies above the curve, the cluster is unlikely to be found at a given survey volume and CL under the \LCDM~paradigm. 
Our new mass estimate is comfortably below the full sky curve. 
When the $2\sigma$ measurement uncertainties are considered, \elgordo~is consistent with the current large-area ACT survey ($13211$ sq. deg; \citealt{Hilton2021}) and can be accommodated even in the initial $1000$ sq. deg survey area \citep{Hasselfield2013}.

Thus, our rarity test with the updated mass shows that \elgordo~is still rare but is compatible with the predictions of \LCDM~cosmology. 
M12 and J14 performed the same rarity test using dynamically inferred and WL masses, respectively, and reached similar conclusions, although they quote lower probabilities because their mass estimates are higher.

\subsubsection{Rarity Test based on the Merging Pair Statistics} \label{rarity_mass_velocity}
As we discussed in \S\ref{merger_scenario}, previous numerical simulation studies assumed an extremely high infall velocity for \elgordo. 
Similarly to masses, mean cluster collision velocities decrease with redshift because the universe at a higher redshift has less time to accrete and accelerate masses. 
Therefore, in principle an extreme collision velocity in a merging pair at high redshift can be viewed as a serious challenge to \LCDM. 
For example, \cite{Lee2010} argued that the high infall velocity $\mytilde3000~$\kms~of the Bullet Cluster is incompatible with \LCDM~(however see \citealt{Bouillot2015}). 
Similarly, \cite{Katz2013} claimed that they did not find any merging pairs like~\elgordo~(with a total mass of $\mytilde2~\times$~\solarmA~and infall velocity of $\mytilde2300$~\kms) in their $1~h^{-3}$ Gpc$^3$ cosmological simulation volume. 
Recently, \cite{Asencio2021} performed a more sophisticated study with a much larger ($216~h^{-3}$ Gpc$^3$) simulation and quoted a $\mytilde6\sigma$ tension with \LCDM.

Three of the most important factors in the \cite{Asencio2021} study are the total mass, mass ratio, and infall velocity of \elgordo. 
With our new mass, mass ratio (\textsection\ref{mass_estimates}), and collision velocity (\textsection\ref{merger_scenario}), we find that the \cite{Asencio2021} method yields no serious tension with \LCDM. 
This is because when we exclude the cases exceeding the escape velocity, the infall velocity at a separation of $\mytilde5.5$~Mpc becomes only $\mytilde450$~\kms~in contrast with the much larger value $\mytilde2500$~\kms~assumed by the authors.

\section{Summary and Conclusion} \label{section_summary}
We have presented an elaborate weak-lensing analysis of the post first-core-crossing merging cluster \elgordo, employing new wide-field \HST/ACS observations in conjunction with archival ACS and WFC3/IR imaging data. 
The new observation covers the $\mytilde3.5\times\mytilde3.5$ Mpc region (corresponding to $\mytilde119$~\sqarcmin) centered on \elgordo, which is $\mytilde2.7$ times larger than our previous WL field. 
The new wide-field data enable us to detect the WL signal beyond the virial radius of the cluster and to determine the cluster mass without extrapolation.
By simultaneously fitting two NFW profiles, we determine the total mass of the cluster to be $M_{200c} = 2.13^{+0.25}_{-0.23}~\times$~\solarmA, which is $\mytilde23$\% less than our previous WL estimate [$M_{200c} = (2.76\pm0.51)~\times$~\solarmA]. 
We verify that this result is in good agreement with our model-independent mass estimate based on aperture mass densitometry. Our updated mass estimate implies that although an \elgordo-like cluster is still rare, its existence does not create a serious tension with \LCDM~even in its parent survey.

Our mass reconstruction confirms the NW-SE binary mass structure and finds a $\mytilde2\sigma$ dissociation between the centroids of the SE mass component and the X-ray cool core. 
The masses of the NW and SE clumps are $M_{200c} = 9.9^{+2.1}_{-2.2}~\times$~\solarm~and $M_{200c} = 6.5^{+1.9}_{-1.4}~\times$~\solarm, respectively. 
This mass ratio is in good agreement with our previous WL result, but contradicts the SL results suggesting that the SE cluster is more massive. We argue that the discrepancy may result from the extrapolation in their SL analysis, as the authors noted. 
When we repeat our analysis with source galaxies only within $\mytilde1.5$ Mpc from each component, our test shows that more mass is indeed associated with the SE cluster at small radii, which indicates that the mass distribution of the SE cluster is more compact than that of the NW cluster. 
Our 2D mass map also shows that the projected density of the SE component is higher at the peak location.

With our updated mass and new physical condition, we revisit the merging scenario of \elgordo. 
The joint study with our updated Monte Carlo simulation and radio-relic constraints favors neither the outgoing nor returning phase. The result is different from the \cite{Ng2015} study, which reports that a returning phase is preferred. We discuss the possibility that the returning case might be supported when we include dynamical friction.
However, more detailed numerical studies are needed in order to explain the peculiar X-ray morphology and current radio relic positions. 
Up to now, existing numerical studies for \elgordo~only considered outgoing phases without utilizing the radio relic data.

Two of the unique properties of \elgordo~are its high mass and large collision velocity at such a high redshift. We discuss the rarity of the system with our improved measurements. We find that our (reduced) mass is more compatible with \LCDM. 
Also, when we impose the condition that the collision velocity cannot exceed the escape velocity based on the timing argument, \elgordo-like mergers can more easily be accommodated within the \LCDM~paradigm.

This work is based on observations made with the NASA/ESA {\it Hubble Space Telescope} and operated by the Association of Universities for Research in Astronomy, Inc. under NASA contract NAS 5-2655. MJJ acknowledges support from the National Research Foundation of Korea under the program nos. 2017R1A2B2004644 and 2017R1A4A1015178. 
MY acknowledge the National Research Foundation of Korea grant funded by the Ministry of Science and ICT, Korea under the program 2019R1C1C1010942. 
JPH acknowledges partial funding for this research from STScI grant number HST-GO-14153.01-A and NSF Astronomy and Astrophysics Grant number AST-1615657.

\software{ 
MultiDrizzle \citep{2002multidrizzle}, 
DrizzlePac \citep{DrizzlePac}, 
SExtractor \citep{Bertin1996}, 
\texttt{CIAO} \citep{CIAO}), 
\texttt{ClusterPyXT} \citep{Alden2019},
{\tt MPFIT} \citep{MPFIT}, 
{\tt FIATMAP} \citep{FIATMAP}, 
{\tt MAXENT} \citep{Jee2007b}, 
MCMAC \citep{MCMAC}
}

\clearpage

\bibliographystyle{aasjournal}


\appendix

\section{Consistency Check for the Mass Reconstruction} \label{mass_map_comparison}
\begin{figure*}
\centering
\includegraphics[width=180mm]{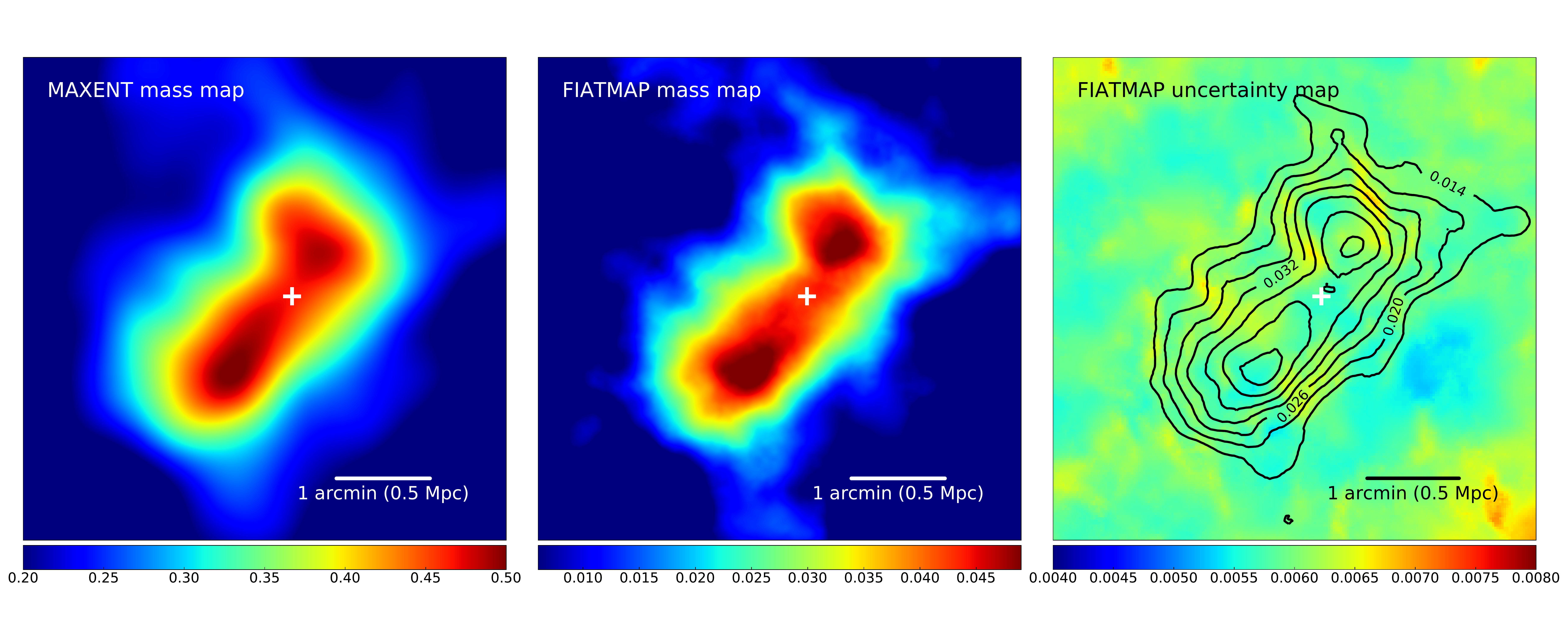}
\caption{Comparison of mass reconstruction of~\elgordo~between 
the {\tt MAXENT} and {\tt FIATMAP} codes. We find that the {\tt MAXENT} version ($left$) is highly consistent with the {\tt FIATMAP} one ($middle$) in terms of the mass substructure.
The scales in $\kappa$ are different because the mass-sheet degeneracy is not lifted. For the {\tt FIATMAP} result, we show the uncertainty (rms) map ($right$) from 1000 bootstrapping runs with the mass contours overlaid.
Note that the peak-to-valley variation of the rms map in the cluster field is extremely small ($\Delta \kappa = 0.0005$, corresponding to $\mytilde0.1\sigma$ of the mass map). 
We indicate the center of mass (white cross) estimated from the best-fit subcluster masses. }
\label{fig:mass_map_comps}
\end{figure*}

We presented mass maps from two different mass reconstruction algorithms: {\tt FIATMAP} \citep{FIATMAP} and {\tt MAXENT} \citep{Jee2007b} in the main text. For fair comparison, here we show the two mass reconstructions side-by-side in Figure~\ref{fig:mass_map_comps} after adjusting the scales. 
We also display our uncertainty map from 1000 bootstrapping of source galaxies in the right panel of Figure~\ref{fig:mass_map_comps}. Note that the variation of the uncertainty across the field is small ($\Delta \kappa = 0.0005$). 
The {\tt FIATMAP} mass map is in good agreement with the {\tt MAXENT} mass map in terms of the mass morphology and the centroid of each substructure.
The {\tt FIATMAP} result is noisier at the cluster outskirt since the algorithm does not employ any regularization. The {\tt MAXENT} code suppresses fluctuations at the field boundary, based on the entropy maximization regularization.

\section{Consistency Check for the Mass Estimation} \label{Ap1}
\subsection{Impact of the prior selection} \label{change_priors}
We repeat the MCMC sampling with various choices of priors in order to investigate the potential impact on the mass estimation. 
Since the NFW halo profile that we adopt is defined by any of the two free parameters from mass $M_{200c}$, concentration $c_{200c}$, and scale radius $r_s$, the possible choices are the combinations of (1) $c_{200c}$ and $M_{200c}$, (2) $c_{200c}$ and $r_s$, and (3) $r_s$ and $M_{200c}$. 
The ranges of the concentration parameter considered are $2 < c < 6$, $1.5 < c < 8$, and $ 0 < c < 10$. 
The first range encloses most of the simulated clusters at high-$z$ (e.g., \citealt{Prada2012}; \citealt{Ludlow2014}; \citealt{DM14}), which is our fiducial choice. 
The last case is the most conservative choice and no galaxy cluster has been found outside this range. 
The second one is the compromise between these two choices. 
As for the mass range, we considered linear and log scale sampling within the range $10^{13}~M_{\sun} < M_{200c} < 10^{16}~M_{\sun}$. 
The linear scale sampling is useful when we desire to impose a uniform prior while the log-scale sampling better matches the mass function of galaxy clusters. 
We set the range of the scale radius to $10\arcsec < r_s < 110\arcsec$, which corresponds to $79~\mbox{kpc} < r_s < 873~\mbox{kpc}$ at $z=0.87$.

\begin{figure}
\centering
\includegraphics[width=55mm]{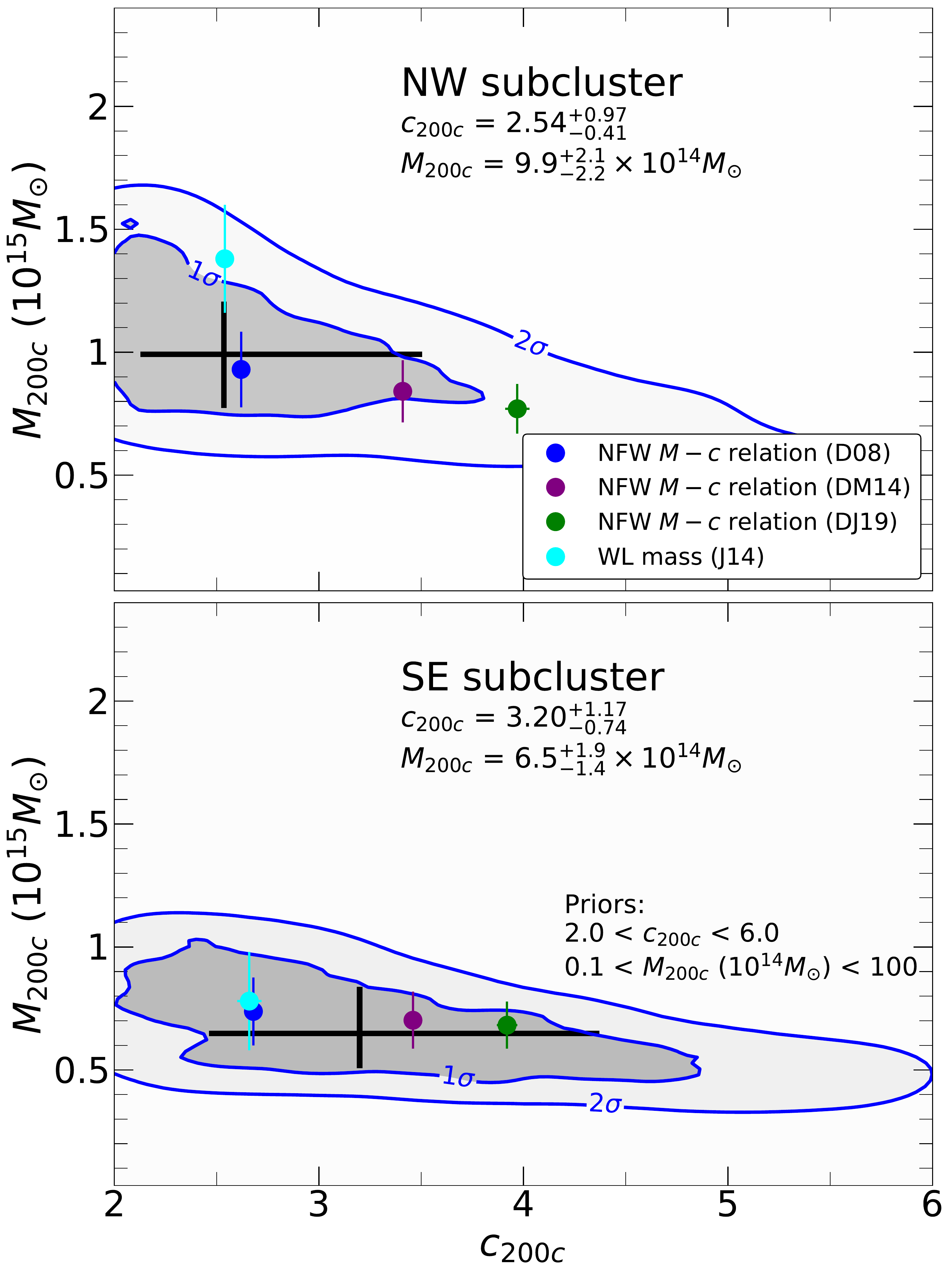}
\includegraphics[width=55mm]{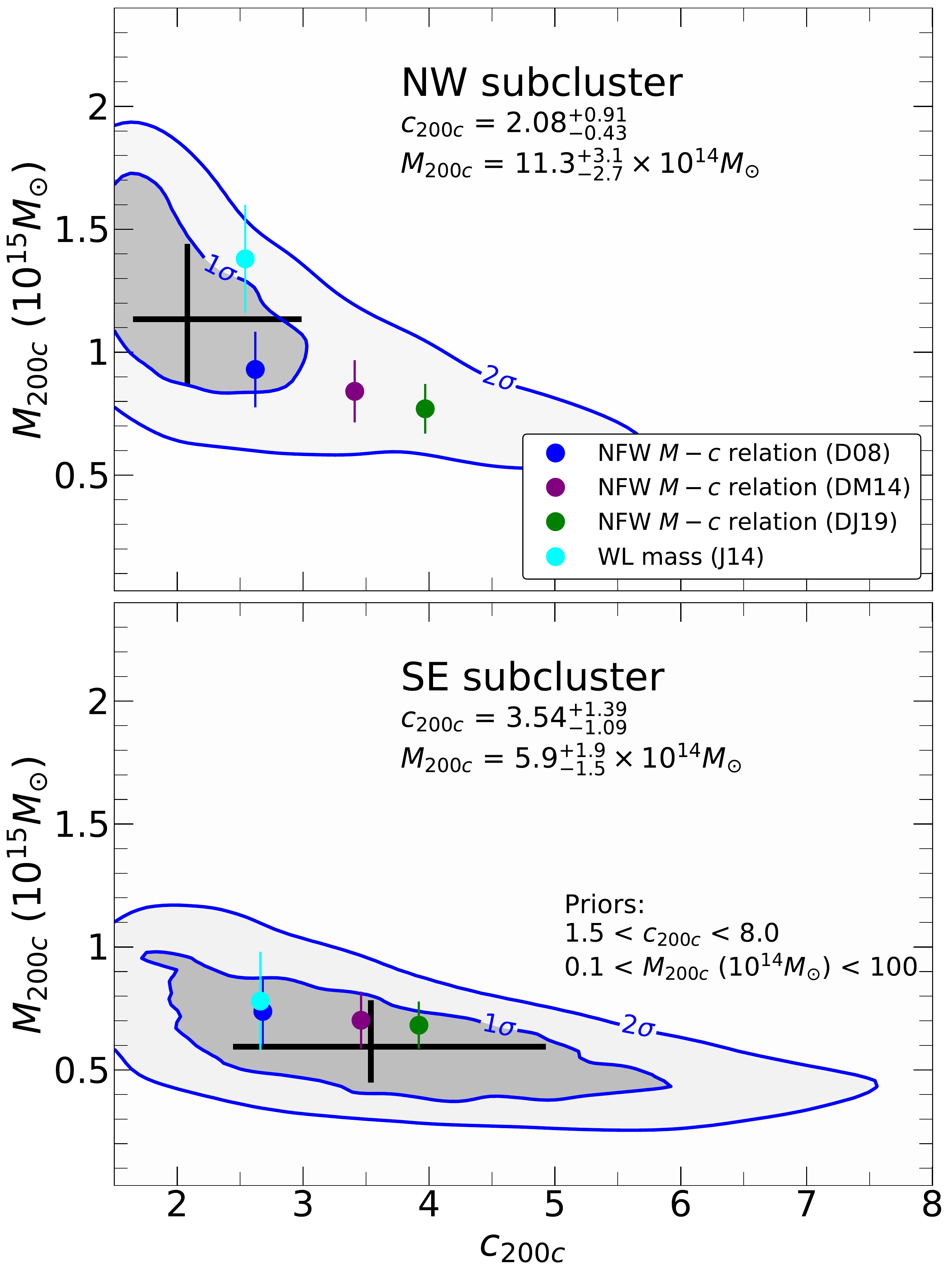}
\includegraphics[width=55mm]{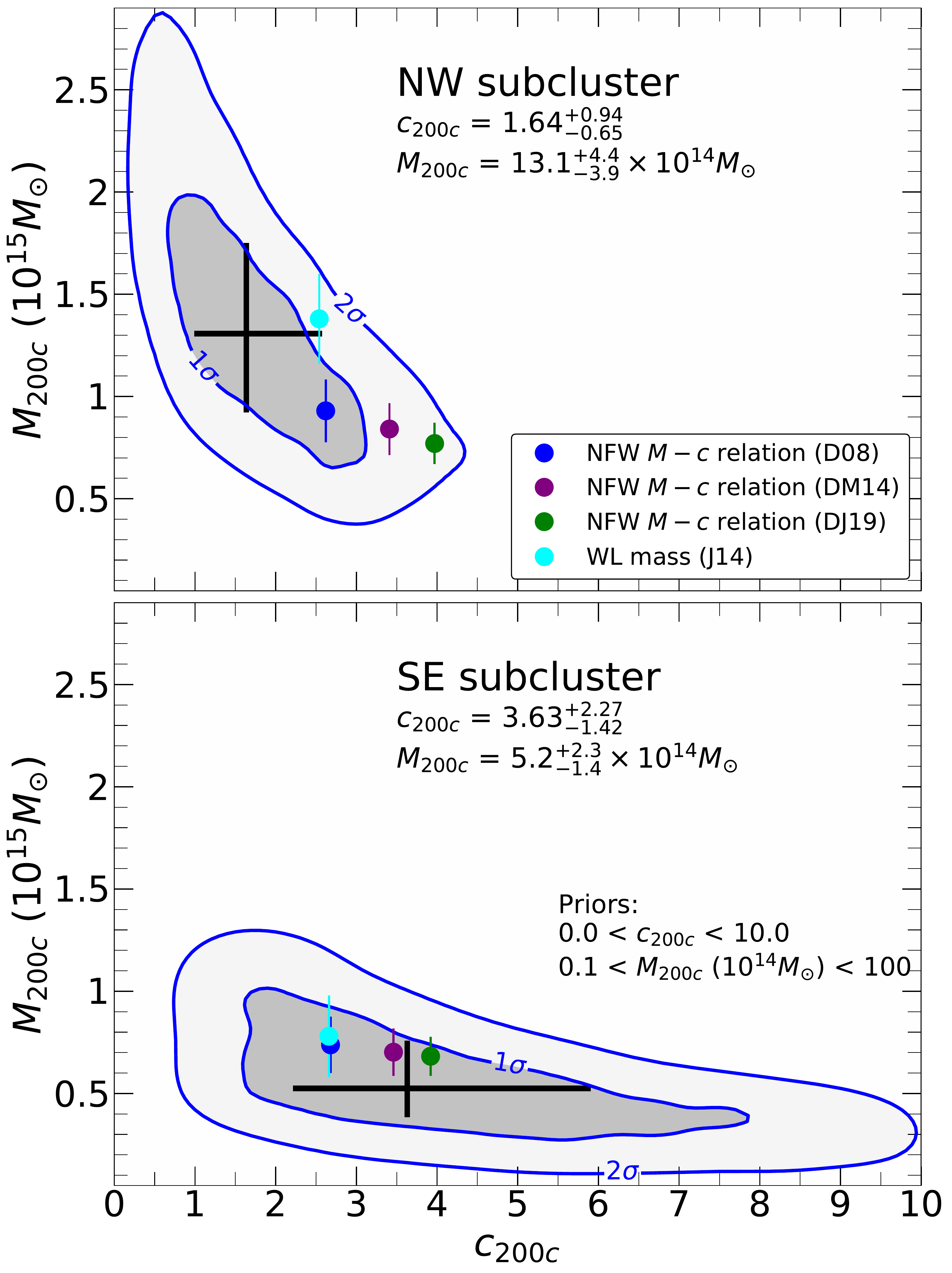}
\includegraphics[width=55mm]{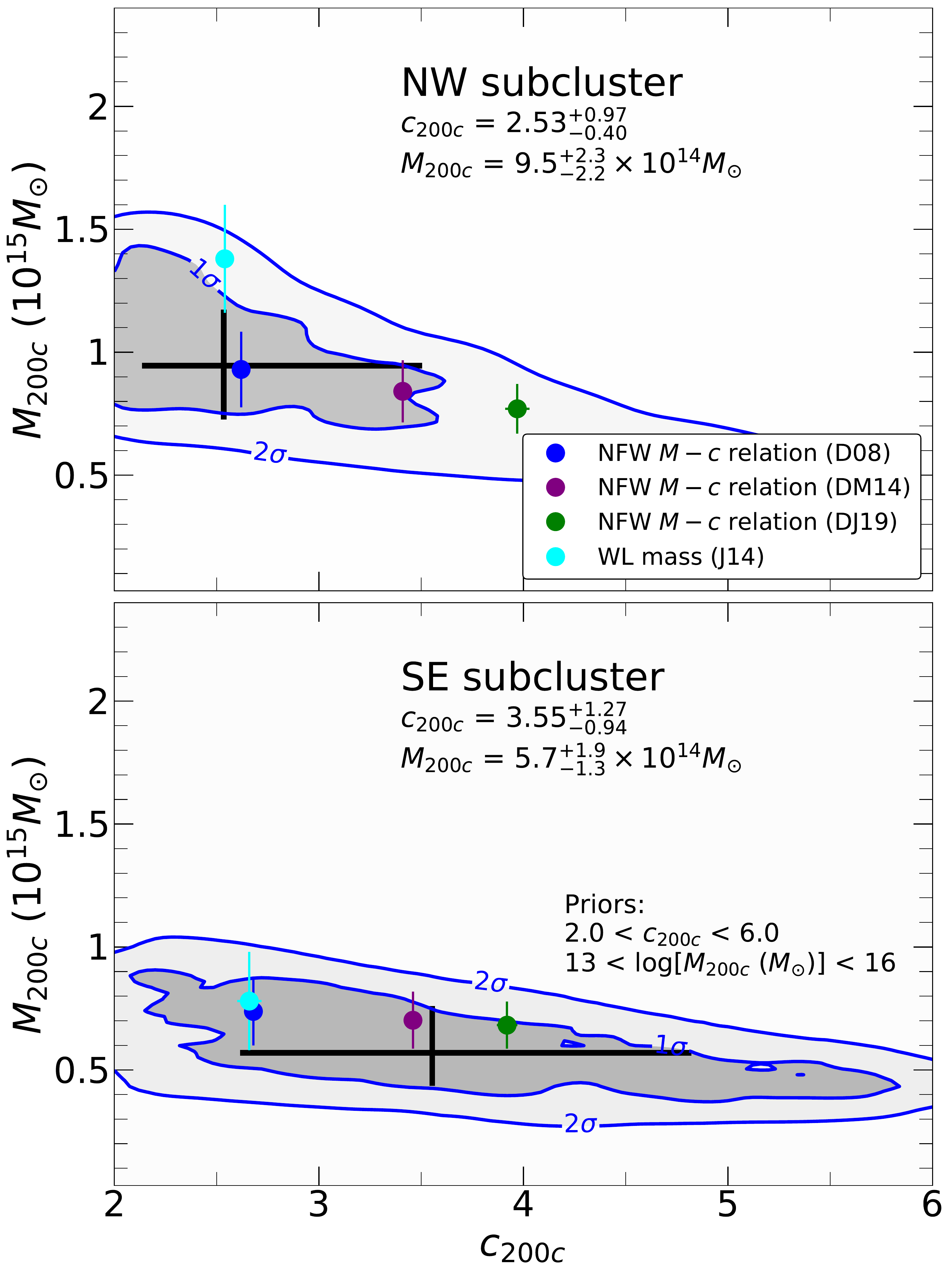}
\includegraphics[width=55mm]{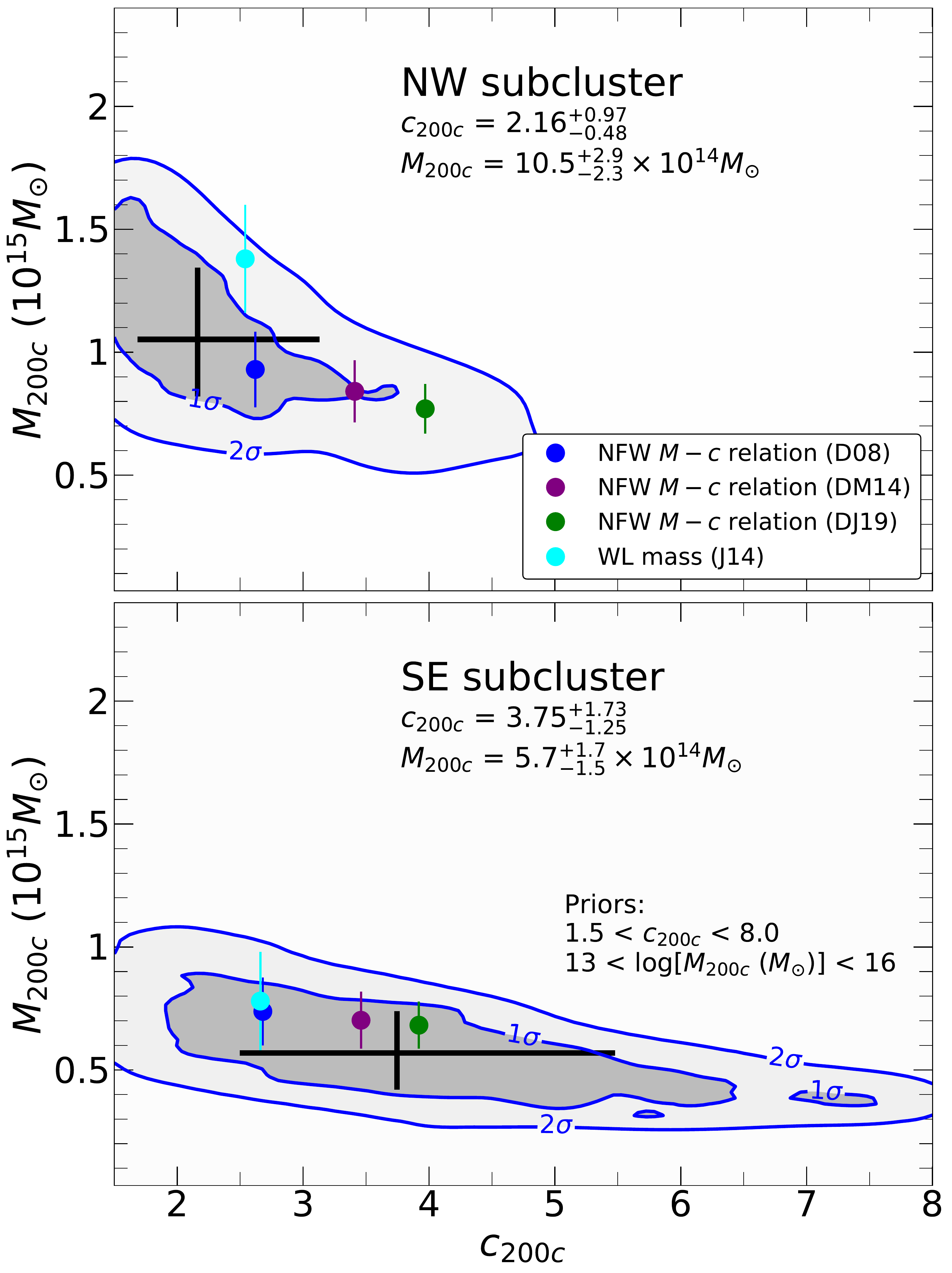}
\includegraphics[width=55mm]{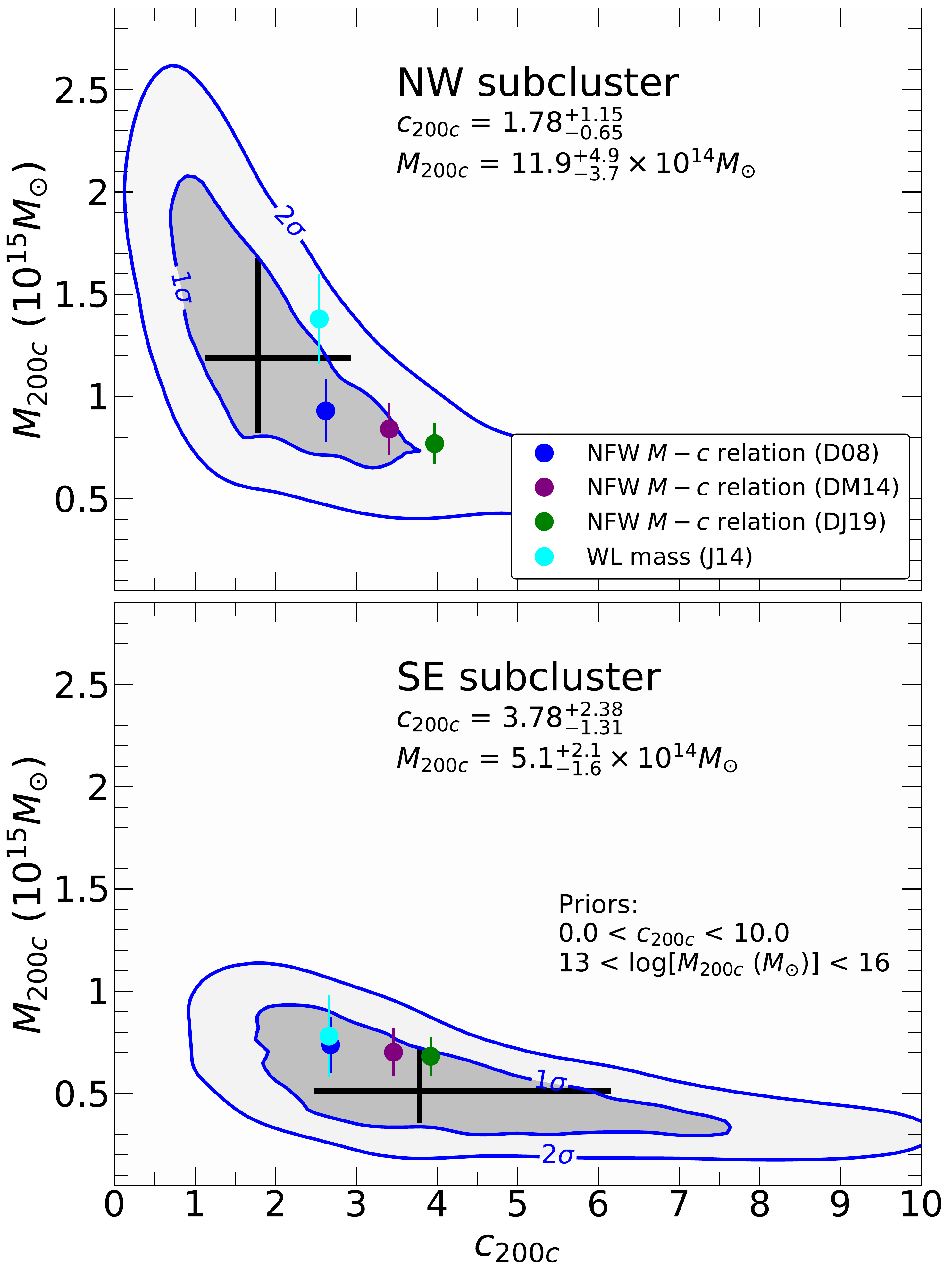}
\caption{Posterior distribution from 100,000 MCMC samples of our two-parameter mass estimation centered at the two mass peaks with different choices of priors. 
The upper two panels adopt the mass prior sampled with a linear-scale (within the range of $10^{13} M_{\odot}$ to $10^{16} M_{\odot}$), while the lower two panels use the log-scaled mass priors ($13 < log (M_{200c}) < 16$). 
The concentration parameter priors are $2 < c_{200c} < 6$, $1.5 < c_{200c} < 8$, and $0 < c_{200c} < 10$ for the left, middle, and right panels, respectively. 
See the description in the caption to Figure~\ref{fig:MCMC_figs} and \textsection\ref{mass_estimates} for more details. }
\label{fig:ap1}
\end{figure}

\begin{figure}
\centering
\includegraphics[width=55mm]{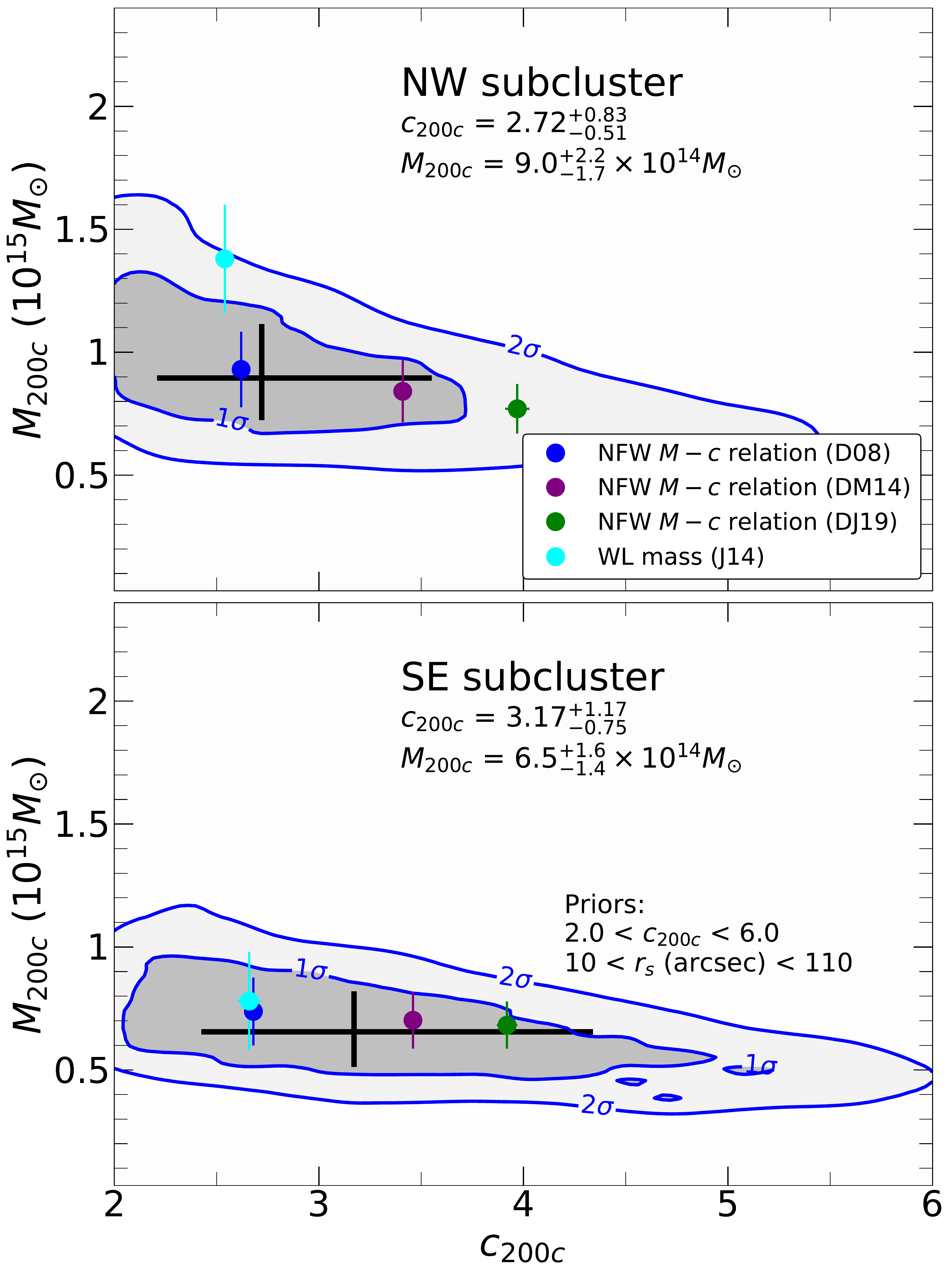}
\includegraphics[width=55mm]{Figure16a_mcmc_c-rs_conc.pdf}
\includegraphics[width=55mm]{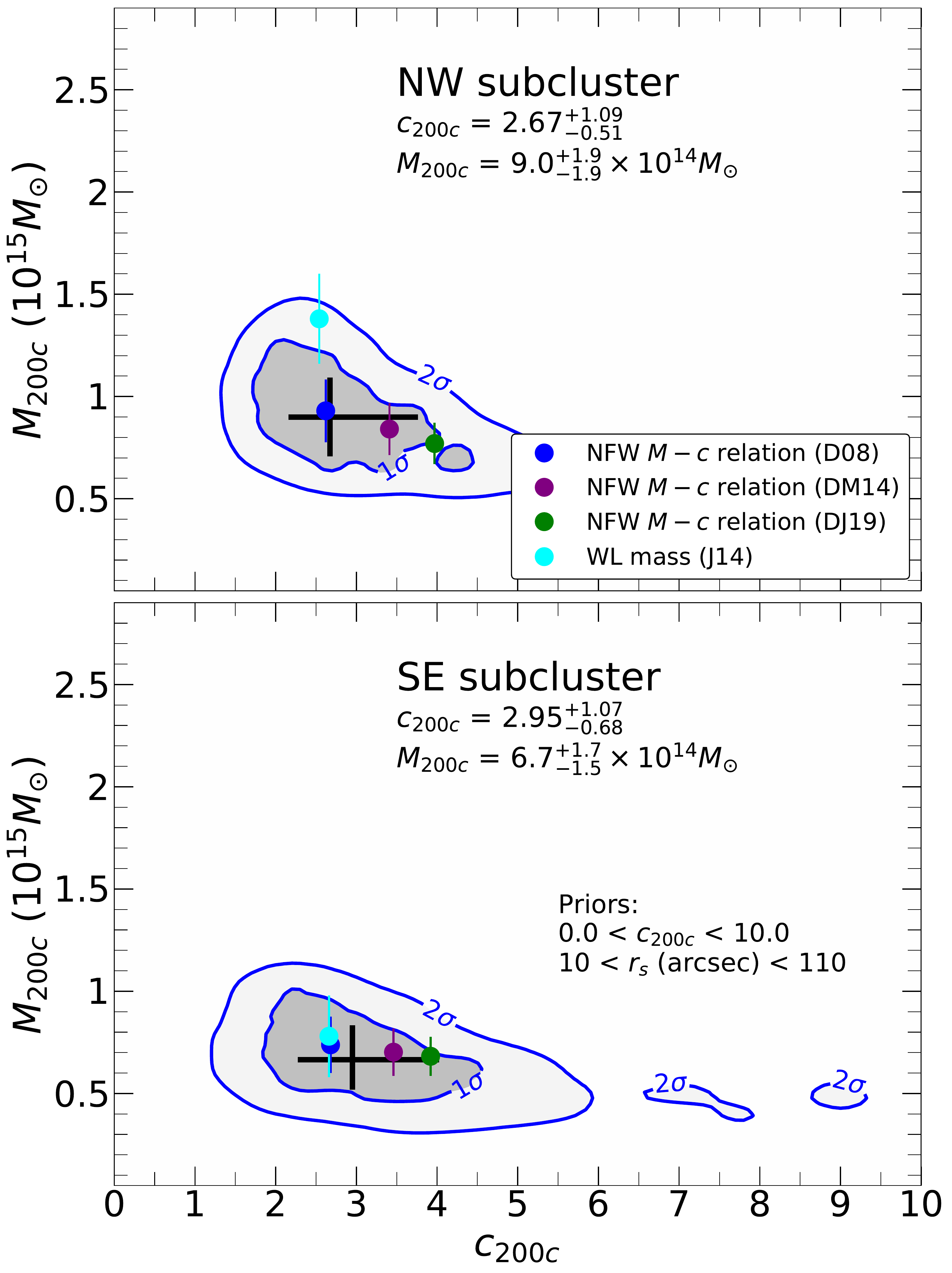}
\includegraphics[width=55mm]{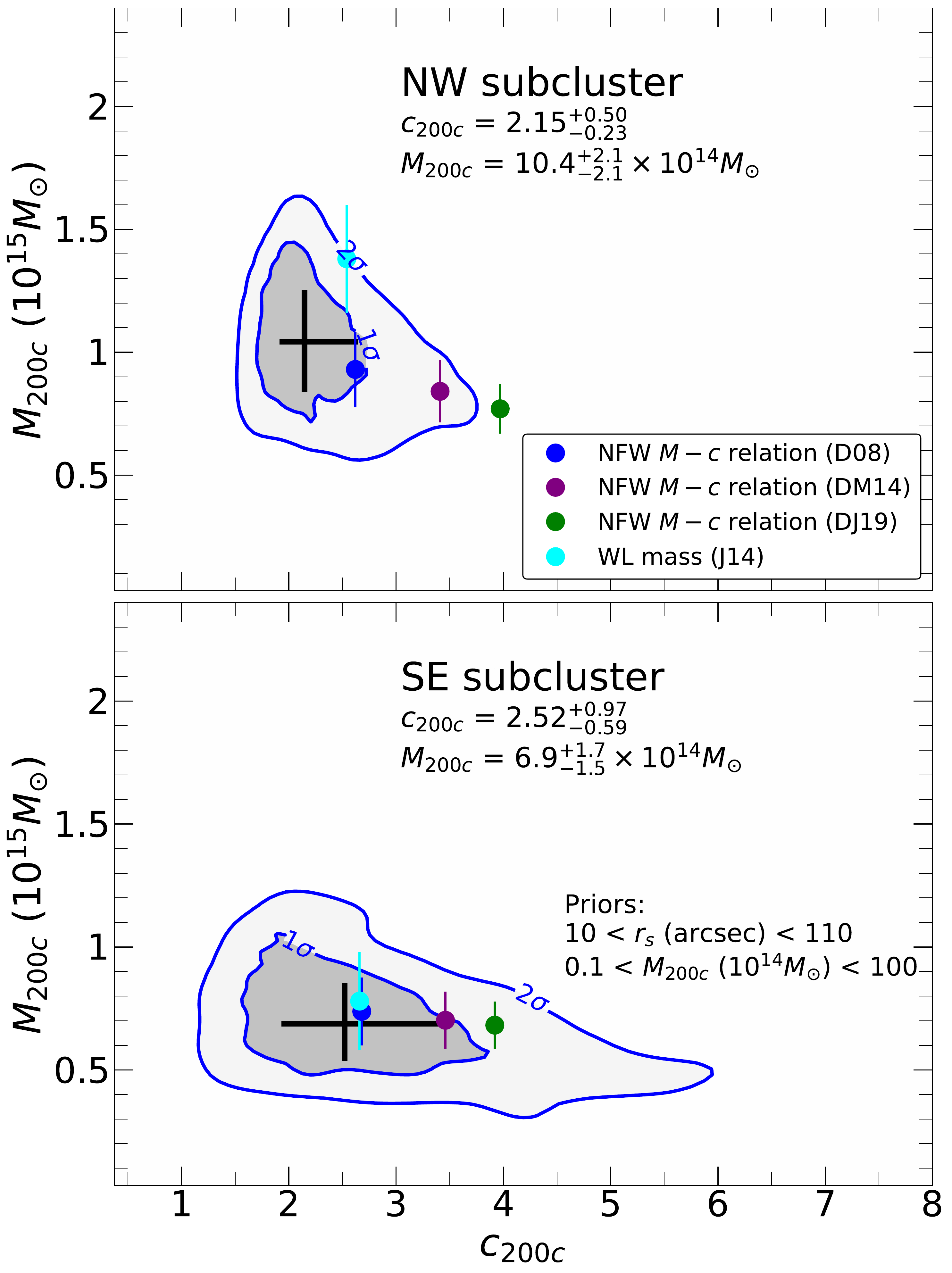}
\includegraphics[width=55mm]{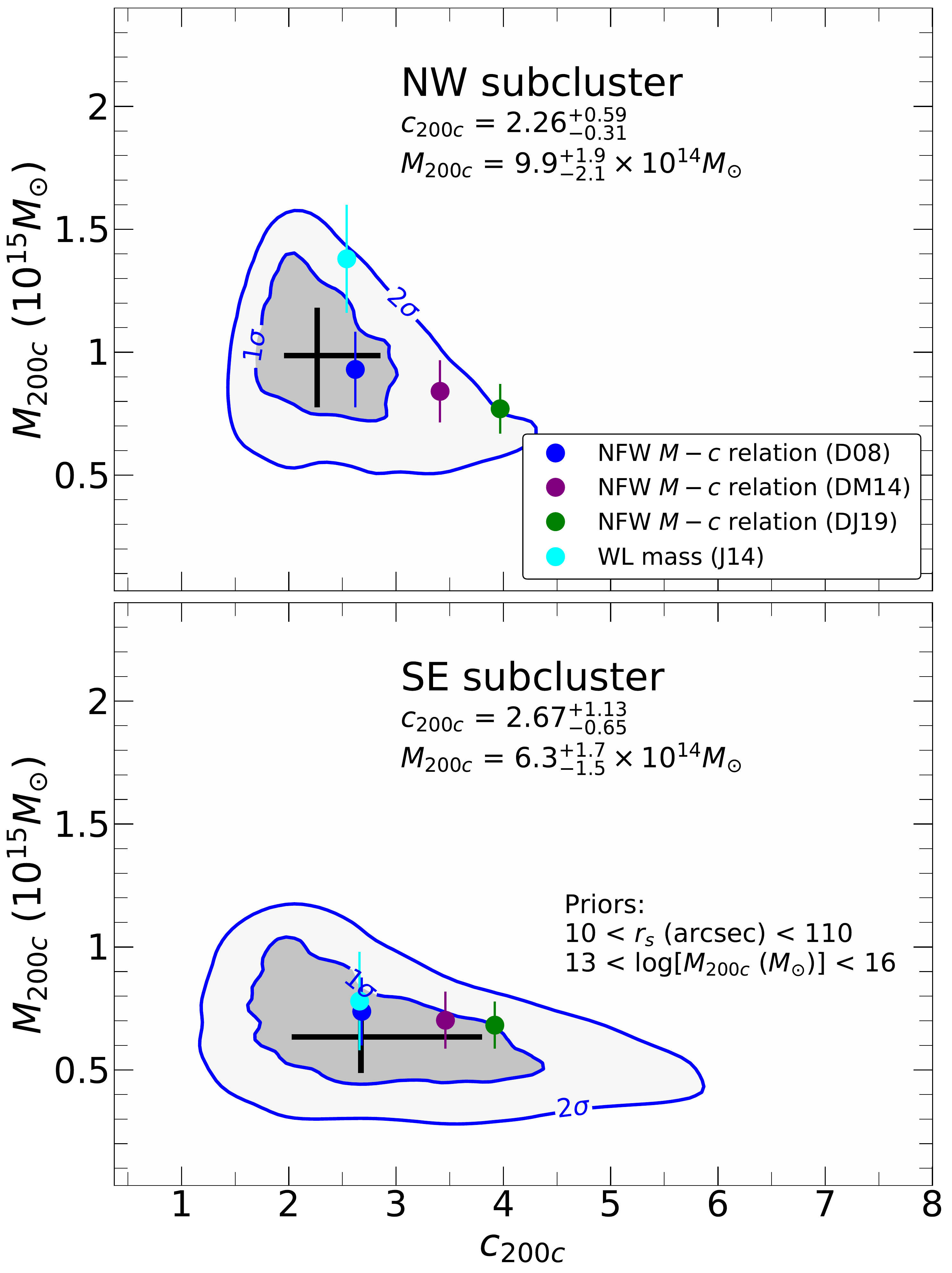}
\caption{Same as Figure~\ref{fig:ap1} with different choices of priors; the scale radius with three cases of concentration (upper) and the scale radius with mass sampled with a linear (lower left) or log (lower right) scale. 
The range of priors for the mass and concentration is identical to that in Figure~\ref{fig:ap1}. 
See the description in the caption to Figure~\ref{fig:MCMC_figs} and \textsection\ref{mass_estimates} for more details. }
\label{fig:ap2}
\end{figure}

Figure~\ref{fig:ap1} and~\ref{fig:ap2} display the posterior distributions from 100,000 MCMC samples for our 11 combinations of priors. 
Our test shows that nearly all results are consistent with one another within their $1\sigma$ uncertainties. 
The largest departure is seen for the concentration parameter prior of $ 0 < c_{200c} < 10$ with a linearly scaled uniform sampling in mass. The mass estimate of the NW component is $\mytilde30$\% higher whereas the SE component is $\mytilde20$\% lower than our fiducial mass estimates. 
Nevertheless, the error bars overlap those of the fiducial results for both components. The mean difference in mass estimate is $\mytilde10$\%. Also, the mass ratio ($M^{SE}_{200c}/M^{NW}_{200c}$) is in good agreement with our fiducial result.

\subsection{Impact of the model selection} \label{check_2Dmodels}
\begin{figure}
\centering
\includegraphics[width=85mm]{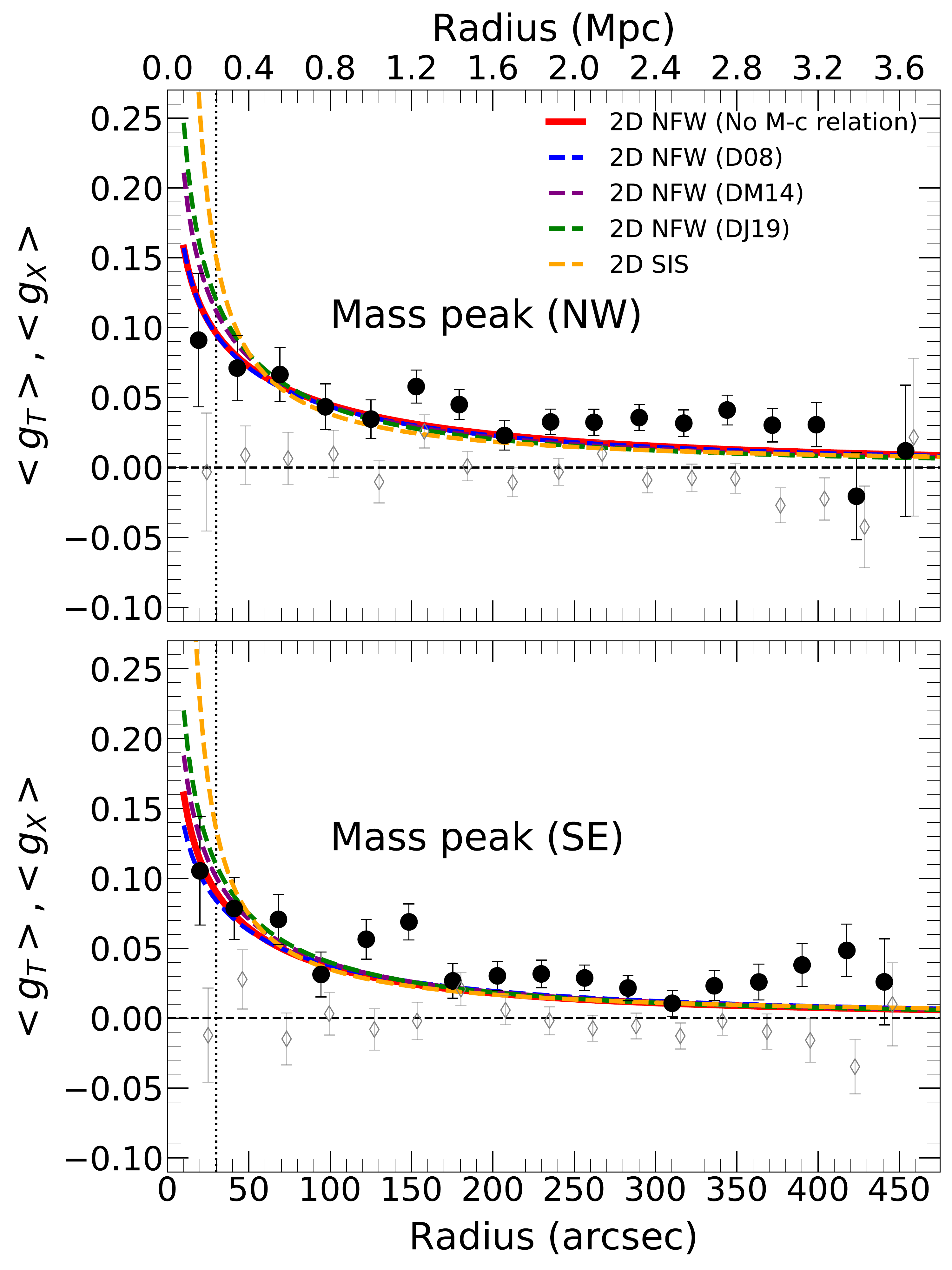}
\caption{Consistency check between 2-halo 2D best-fit models and tangential shears. 
Reduced tangential shear profiles are measured with fixed centers at two mass peaks. 
The tangential shear ($g_{T}$) and the cross shear ($g_{X}$, the 45\degr~rotation of the source images) are indicated by filled circles and open diamonds, respectively. 
We exclude the sources at $r < 30$\arcsec~(vertical dotted lines, see text) for model fitting. 
We show the best-fit NFW and singular isothermal sphere (SIS, orange) model halos. 
For the best-fit NFW model, we determine the model with and without the mass-concentration ($M-c$) relations. 
When determining the best-fit 2D NFW model without $M-c$ relations (red), we estimate the two free parameters of the NFW profile simultaneously via the MCMC method (see text). 
We adopt three different $M-c$ relations from \cite{Duffy08} (blue), \cite{DM14} (purple), and \cite{DJ19} (green). Note that the best-fit models are consistent at radius $r > 50$\arcsec. }
\label{fig:shear_profiles}
\end{figure}

For a consistency check, we plot the best-fit 2D SIS and NFW models with and without $M-c$ relations for both components with the tangential shear profiles in Figure~\ref{fig:shear_profiles}. 
The tangential shears defined by Equation~\ref{tan_shear} of the source galaxies are azimuthally averaged and plotted. 
When compared with our fiducial results, that is the 2D NFW fitting without $M-c$ relation, the models are consistent at $r > 50$\arcsec~($\mytilde400$ kpc). 
The best-fit models show a good consistency with the tangential shears in general, although for each component there are two significant enhancements at large radii in the observed profiles. 
The inner bump between $r\sim100$\arcsec~and $r\sim150$\arcsec~can be attributed to the contribution from the other component. 
The outer bump may originate from either the large-scale structure around the cluster or be due to an effect of field boundary (the annulus used for the tangential shear can complete a circle only up to $r\sim300$\arcsec~when we place the center at the NW or SE clump).

\section{The Mass Estimation of Full System} \label{mass_total}
The total mass of \elgordo~is obtained using the following procedure; 
we first construct two NFW profiles defined by the mass and concentration parameters of two halos from each MCMC sample and populate these halo profiles into a three-dimensional ($500^3$) grid. 
We then obtain a cumulative NFW density distribution centered at the center of mass and calculate the corresponding mass. 
We repeat this process for 100,000 MCMC samples to obtain the uncertainty of the mass of the full system. 
We assume that the two halos are at the same distance from us. This assumption is justified by the fact that the the line-of-sight velocity difference between the two components is small. 
This implies that the angle between the plane of the sky and the merger axis is small, which is supported by the Monte Carlo simulation (e.g., $\mytilde21$\degr; \citealt{Ng2015}). 
J14 also showed that the combined mass of the cluster is not significantly affected unless the merger axis is greater than 60\degr.

\begin{figure}
\centering
\includegraphics[width=85mm]{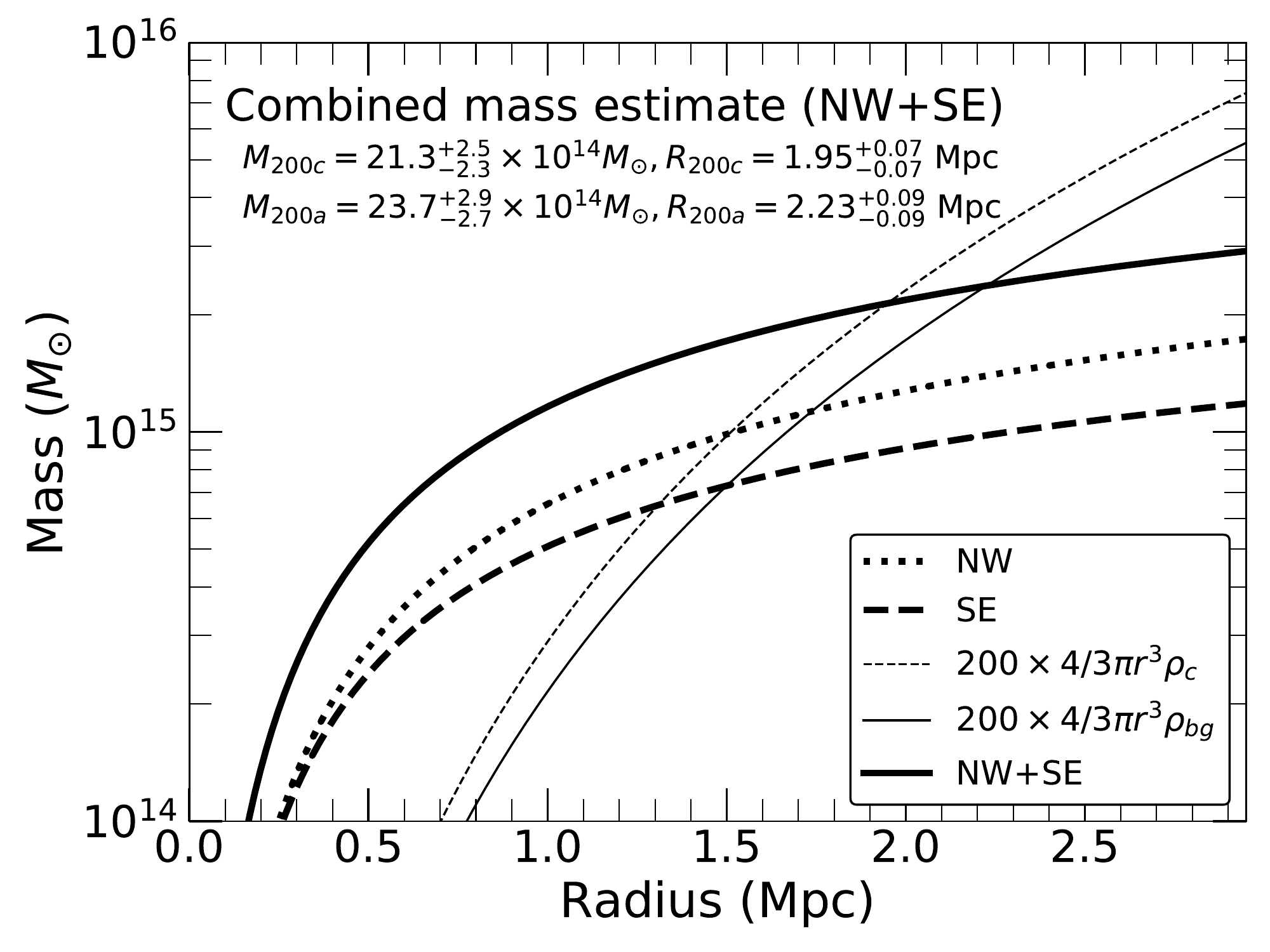}
\caption{Total mass of the halo within a given radius. 
We use the 2D NFW fitting results without the $M-c$ relation. 
The masses within spherical volumes for individual components (NW and SE) and the sum of the two are presented by thick dashed, dotted and solid lines, respectively. 
We assume that the projected distance between two components ($\mytilde770$~kpc) is the actual physical separation (i.e., the angle between the merger axis and the plane of the sky is zero) and choose the center of mass of the two components for the origin. 
The thin dashed and solid lines are the mass within the sphere spherical volume where the density inside becomes 200 times the critical and mean densities of the universe at the cluster redshift, respectively. 
We determine the $M_{200c}$ and $M_{200a}$ values by locating where the thin and thick lines intersect. }
\label{fig:combined_mass_profile}
\end{figure}

Figure~\ref{fig:combined_mass_profile} shows the resulting mass profiles of the two individual subclusters (thick dashed and dotted lines) centered at each component and the sum of them measured from the center of mass (thick solid line). 
We plot the cumulative mass within a sphere at radius $r$ where the density inside is equal to 200 times the critical (mean) density of the universe as thin dashed (thin solid) line. 
The $M_{200c}$ ($M_{200a}$) value is determined by finding the intersection point between the combined mass profile (thick solid) and thin dashed (thin solid) lines.

\clearpage

\end{document}